\documentclass[prd,aps,twocolumn,superscriptaddress,nofootinbib,letterpaper,balancelastpage,notitlepage,floatfix]{revtex4-2}

\usepackage{amsmath,graphicx,epsfig}
\usepackage{epstopdf}
\usepackage{euscript}
\usepackage{amsfonts}
\usepackage{amssymb}
\usepackage{float}
\usepackage[colorlinks = true,
            linkcolor = blue,
            urlcolor  = blue,
            citecolor = blue,
            anchorcolor = blue]{hyperref}

\def\prn#1{{\left(#1\right)}}

\def\sbrk#1{{\left[#1\right]}}

\def\bra#1{{\langle#1|}}
\def\abs#1{{\left|#1\right|}}

\def\pdbyd#1#2{{\frac{\partial#1}{\partial#2}}}

\def\cg(#1,#2)(#3,#4)(#5,#6){\bra{#1,#2,#3,#4}#5,#6\rangle}

\def\ts#1{{_{\mbox{\scriptsize #1}}}}

\def\threej(#1,#2)(#3,#4)(#5,#6){\begin{pmatrix}#1&#3&#5\\#2&#4&#6\end{pmatrix}}
\def\sixj(#1,#2,#3)(#4,#5,#6){\begin{Bmatrix}#1&#2&#3\\#4&#5&#6\end{Bmatrix}}
\def\ninej(#1,#2,#3)(#4,#5,#6)(#7,#8,#9){\begin{Bmatrix}#1&#2&#3\\#4&#5&#6\\#7&#8&#9\end{Bmatrix}}

\DeclareMathOperator{\sech}{sech}

\def\bs{\boldsymbol}
\def\mc{\mathcal}

\usepackage[arrows]{ezedits}
\defineEdit{DK}{\color[rgb]{1.0,0.0,0.0}}{\color[rgb]{1.0,0.0,0.0}}
\defineEdit{EK}{\color[rgb]{,0.50,0.50}}{\color[rgb]{0,0.50,0.50}}
\defineEdit{MP}{\color[rgb]{0.255,0,0.255}}{\color[rgb]{0.255,0,0.255}}
\defineEdit{GL}{\color[rgb]{0,0,0.255}}{\color[rgb]{0,0,0.255}}
\defineEdit{DB}{\color[rgb]{1.0,0.5,0.0}}{\color[rgb]{1.0,0.5,0.0}}
\defineEdit{AW}{\color[rgb]{0,0.25,1}}{\color[rgb]{0,0.5,1}}
\defineEdit{JES}{\color[rgb]{0,1,0.2}}{\color[rgb]{0,1,0.2}}
\defineEdit{TS}{\color[rgb]{0.63,0.12,0.94}}{\color[rgb]{0.63,0.12,0.94}}
\defineEdit{IAS}{\color[rgb]{0,0.7,0.5}}{\color[rgb]{0,0.7,0.5}}
\defineEdit{SP}{\color[rgb]{0,1,0.45}}{\color[rgb]{0,1,0.15}}
\defineEdit{JoSm}{\color[rgb]{0.2,0.5,0.}}{\color[rgb]{0,1,0.15}}
\defineEdit{AK}{\color[rgb]{1.0,0.0,0.0}}{\color[rgb]{1.0,0.0,0.0}}

\begin{document}

\title{What can a GNOME do?\\ ~~ \\Search targets for the Global Network of \\Optical Magnetometers for Exotic physics searches} %==================================

%\author{GNOME collaboration}

\author{S. Afach}
\affiliation{Helmholtz-Institut, GSI Helmholtzzentrum fur Schwerionenforschung, Mainz 55128, Germany}
\affiliation{Johannes Gutenberg University, Mainz 55128, Germany}

\author{D. Aybas Tumturk}
\affiliation{Department of Physics, University of California at Berkeley, Berkeley, California 94720-7300, USA}

\author{H. Bekker}
\affiliation{Helmholtz-Institut, GSI Helmholtzzentrum fur Schwerionenforschung, Mainz 55128, Germany}
\affiliation{Johannes Gutenberg University, Mainz 55128, Germany}

\author{B. C. Buchler}
\affiliation{Research School of Physics, Australian National University, Canberra ACT 2601, Australia}

\author{D. Budker}
\affiliation{Johannes Gutenberg University, Mainz 55128, Germany}
\affiliation{Helmholtz-Institut, GSI Helmholtzzentrum fur Schwerionenforschung, Mainz 55128, Germany}
\affiliation{Department of Physics, University of California at Berkeley, Berkeley, California 94720-7300, USA}

\author{K. Cervantes}
\affiliation{Department of Physics, California State University -- East Bay, Hayward, California 94542-3084, USA}

\author{A. Derevianko}
\affiliation{Department of Physics, University of Nevada, Reno, Nevada 89557, USA}

\author{J. Eby}
\affiliation{Kavli Institute for the Physics and Mathematics of the Universe (WPI), The University of Tokyo Institutes for Advanced Study, The University of Tokyo, Kashiwa, Chiba 277-8583, Japan}

\author{N. L. Figueroa}
\affiliation{Johannes Gutenberg University, Mainz 55128, Germany}
\affiliation{Helmholtz-Institut, GSI Helmholtzzentrum fur Schwerionenforschung, Mainz 55128, Germany}

\author{R. Folman}
\affiliation{Department of Physics, Ben-Gurion University of the Negev, Be’er Sheva 84105, Israel}

\author{D. Gavilán Martín}
\affiliation{Johannes Gutenberg University, Mainz 55128, Germany}
\affiliation{Helmholtz-Institut, GSI Helmholtzzentrum fur Schwerionenforschung, Mainz 55128, Germany}

\author{M. Givon}
\affiliation{Department of Physics, Ben-Gurion University of the Negev, Be’er Sheva 84105, Israel}

\author{Z. D. Grujić}
\affiliation{Physics Department, University of Fribourg, CH-1700 Fribourg, Switzerland}
\affiliation{Institute of Physics Belgrade, Pregrecica 118, Belgrade 11080, Serbia}

\author{H. Guo}
\affiliation{State Key Laboratory of Advanced Optical Communication Systems and Networks, Department of Electronics, and Center for Quantum Information Technology, Peking University, Beijing 100871, China }

\author{P. Hamilton}
\affiliation{Department of Physics \& Astronomy, University of California at Los Angeles, Los Angeles, California, 90095-1547, USA}

\author{M. P. Hedges}
\affiliation{Research School of Physics, Australian National University, Canberra ACT 2601, Australia}

\author{D. F. Jackson Kimball}
\email{derek.jacksonkimball@csueastbay.edu}
\affiliation{Department of Physics, California State University -- East Bay, Hayward, California 94542-3084, USA}

\author{S. Khamis}
\affiliation{Department of Physics \& Astronomy, University of California at Los Angeles, Los Angeles, California, 90095-1547, USA}

\author{D. Kim}
\thanks{present address: Mechatronics Research, Samsung Electronics, Hwaseong, 18448, South Korea}
\affiliation{Department of Physics, KAIST, Daejeon 34141, South Korea}
\affiliation{Center for Axion and Precision Physics Research, IBS, Daejeon, 34051, South Korea}

\author{E. Klinger}
\thanks{present address: Université de Franche-Comté, SupMicroTech-ENSMM, UMR 6174 CNRS, institut FEMTO-ST, 25000 Besançon, France}
\affiliation{Helmholtz-Institut, GSI Helmholtzzentrum fur Schwerionenforschung, Mainz 55128, Germany}
\affiliation{Johannes Gutenberg University, Mainz 55128, Germany}

\author{A. Kryemadhi}
\affiliation{Dept. Computing,Math \& Physics, Messiah University, Mechanicsburg PA 17055, USA}

\author{X. Liu}
\affiliation{State Key Laboratory of Advanced Optical Communication Systems and Networks, Department of Electronics, and Center for Quantum Information Technology, Peking University, Beijing 100871, China }

\author{G. Łukasiewicz}
\affiliation{Marian Smoluchowski Institute of Physics, Jagiellonian University,  Łojasiewicza 11, 30-348, Kraków, Poland}

\author{H. Masia-Roig}
\affiliation{Johannes Gutenberg University, Mainz 55128, Germany}
\affiliation{Helmholtz-Institut, GSI Helmholtzzentrum fur Schwerionenforschung, Mainz 55128, Germany}

\author{M. Padniuk}
\affiliation{Marian Smoluchowski Institute of Physics, Jagiellonian University,  Łojasiewicza 11, 30-348, Kraków, Poland}

\author{C. A. Palm}
\affiliation{Department of Physics, California State University -- East Bay, Hayward, California 94542-3084, USA}

\author{S. Y. Park}
\thanks{present address: Department of Physics, University of Colorado, Boulder Colorado 80309, USA }
\affiliation{Department of Physics \& Astronomy, Oberlin College, Oberlin, Ohio 44074, USA}

\author{H. R. Pearson}
\affiliation{Department of Physics \& Astronomy, Oberlin College, Oberlin, Ohio 44074, USA}

\author{X. Peng}
\affiliation{State Key Laboratory of Advanced Optical Communication Systems and Networks, Department of Electronics, and Center for Quantum Information Technology, Peking University, Beijing 100871, China }

\author{M. Pospelov}
\affiliation{School of Physics and Astronomy, University of Minnesota, Minneapolis, MN 55455, USA}
\affiliation{William I. Fine Theoretical Physics Institute, School of Physics and Astronomy, University of
Minnesota, Minneapolis, MN 55455, USA}

\author{S. Pustelny}
\affiliation{Marian Smoluchowski Institute of Physics, Jagiellonian University,  Łojasiewicza 11, 30-348, Kraków, Poland}

\author{Y. Rosenzweig}
\affiliation{Department of Physics, Ben-Gurion University of the Negev, Be’er Sheva 84105, Israel}

\author{O. M. Ruimi}
\affiliation{Racah Institute of Physics, Hebrew University of Jerusalem, Jerusalem 9190401, Israel}

\author{T.~Scholtes}
\affiliation{Physics Department, University of Fribourg, CH-1700 Fribourg, Switzerland}
\affiliation{Leibniz Institute of Photonic Technology, Albert-Einstein-Strasse 9, D-07745 Jena, Germany}

\author{P. C. Segura}
\thanks{present address: Department of Physics, Harvard University, Cambridge, Massachusetts, 02138, USA}
\affiliation{Department of Physics \& Astronomy, Oberlin College, Oberlin, Ohio 44074, USA}

\author{Y. K. Semertzidis}
\affiliation{Center for Axion and Precision Physics Research, IBS, Daejeon, 34051, South Korea}
\affiliation{Department of Physics, KAIST, Daejeon 34141, South Korea}

\author{Y. C. Shin}
\affiliation{Center for Axion and Precision Physics Research, IBS, Daejeon, 34051, South Korea}

\author{J. A. Smiga}
\thanks{present address: Department of Physics and Astronomy, University of Rochester, Rochester, New York 14627, USA}
\affiliation{Johannes Gutenberg University, Mainz 55128, Germany}
\affiliation{Helmholtz-Institut, GSI Helmholtzzentrum fur Schwerionenforschung, Mainz 55128, Germany}

\author{Y.~V.~Stadnik}
\affiliation{School of Physics, University of Sydney, Sydney, NSW 2006, Australia}

\author{J. E. Stalnaker}
\affiliation{Department of Physics \& Astronomy, Oberlin College, Oberlin, Ohio 44074, USA}

\author{I. A. Sulai}
\affiliation{Department of Physics \& Astronomy, Bucknell University, Lewisburg, Pennsylvania 17837, USA}

\author{D. Tandon}
\affiliation{Department of Physics \& Astronomy, Oberlin College, Oberlin, Ohio 44074, USA}

\author{K. Vu}
\affiliation{Department of Physics, California State University -- East Bay, Hayward, California 94542-3084, USA}

\author{A.~Weis}
\affiliation{Physics Department, University of Fribourg, CH-1700 Fribourg, Switzerland}

\author{A. Wickenbrock}
\affiliation{Johannes Gutenberg University, Mainz 55128, Germany}
\affiliation{Helmholtz-Institut, GSI Helmholtzzentrum fur Schwerionenforschung, Mainz 55128, Germany}

\author{T. Z. Wilson}
\thanks{present address: Department of Physics, University of Illinois at Urbana-Champaign, Urbana, Illinois 61801, USA}
\affiliation{Department of Physics, California State University -- East Bay, Hayward, California 94542-3084, USA}

\author{T. Wu}
\affiliation{State Key Laboratory of Advanced Optical Communication Systems and Networks, Department of Electronics, and Center for Quantum Information Technology, Peking University, Beijing 100871, China }

\author{W. Xiao}
\affiliation{State Key Laboratory of Advanced Optical Communication Systems and Networks, Department of Electronics, and Center for Quantum Information Technology, Peking University, Beijing 100871, China }

\author{Y. Yang}
\affiliation{State Key Laboratory of Advanced Optical Communication Systems and Networks, Department of Electronics, and Center for Quantum Information Technology, Peking University, Beijing 100871, China }
\affiliation{Beijing Institute of Applied Meteorology, Beijing 100029, China}

\author{D. Yu}
\affiliation{State Key Laboratory of Advanced Optical Communication Systems and Networks, Department of Electronics, and Center for Quantum Information Technology, Peking University, Beijing 100871, China }

\author{F. Yu}
\affiliation{PRISMA+ Cluster of Excellence \& Mainz Institute for Theoretical Physics, Johannes Gutenberg University, Mainz 55128, Germany}

\author{J. Zhang}
\affiliation{State Key Laboratory of Advanced Optical Communication Systems and Networks, Department of Electronics, and Center for Quantum Information Technology, Peking University, Beijing 100871, China }

\author{Y. Zhao}
\affiliation{State Key Laboratory of Advanced Optical Communication Systems and Networks, Department of Electronics, and Center for Quantum Information Technology, Peking University, Beijing 100871, China }

\date{\today}

%%%----------------------------------------------------------------------

%\doublespacing

\begin{abstract}
Numerous observations suggest that there exist undiscovered beyond-the-Standard-Model particles and fields.
Because of their unknown nature, these exotic particles and fields could interact with Standard Model particles in many different ways and assume a variety of possible configurations.
Here we present an overview of the Global Network of Optical Magnetometers for Exotic physics searches (GNOME), our ongoing experimental program designed to test a wide range of exotic physics scenarios.
The GNOME experiment utilizes a worldwide network of shielded atomic magnetometers (and, more recently, comagnetometers) to search for spatially and temporally correlated signals due to torques on atomic spins from exotic fields of astrophysical origin.
We survey the temporal characteristics of a variety of possible signals currently under investigation such as those from topological defect dark matter (axion-like particle domain walls), axion-like particle stars, solitons of complex-valued scalar fields (Q-balls), stochastic fluctuations of bosonic dark matter fields, a solar axion-like particle halo, and bursts of ultralight bosonic fields produced by cataclysmic astrophysical events such as binary black hole mergers.
\end{abstract}
%\pacs{PACS. ???}

%%%----------------------------------------------------------------------

%\doublespacing

\maketitle

%%%%%%%%%%%%%%%%%%%%%%%%%%%%%%%%%%%%%%%%%%%%%%%
%%%%%%%%%%%%%%%%%%%%%%%%%%%%%%%%%%%%%%%%%%%%%%%
\section{Introduction}
%%%%%%%%%%%%%%%%%%%%%%%%%%%%%%%%%%%%%%%%%%%%%%%
%%%%%%%%%%%%%%%%%%%%%%%%%%%%%%%%%%%%%%%%%%%%%%%

There are widespread hints from nature suggesting there exist exotic, heretofore undiscovered particles.
Perhaps the most prominent hint is the accumulated evidence for dark matter.
A leading hypotheses to explain dark matter is that it consists of ultralight bosons such as axions or axion-like particles (ALPs) with masses $m_a \ll 1\,{\rm{eV}}$ \cite{kimball2022search,graham2015experimental}.
Such ultralight bosonic dark matter (UBDM) can couple to Standard Model particles through a variety of ``portals'' \cite{graham2016dark,safronova2018search}, one of which is the direct interaction of the UBDM field with atomic spins \cite{stadnik2014axion,graham2018spin}.
If such an interaction exists, a UBDM field would generate a spin-dependent energy shift similar to that caused by the Zeeman effect due to an external magnetic field.
This opens the possibility of using atomic-spin-based magnetometers \cite{Bud13,budker2007optical} to search for UBDM.

Several experiments use atomic magnetometers and nuclear magnetic resonance (NMR) techniques to search for the interaction of UBDM fields with spins \cite{budker2014proposal,abel2017search,Garcon2017,wu2019search,garcon2019constraints,Aybas2021,roussy2021experimental,Ayb21CASPErE,jiang2021search,bloch2022new,lee2022laboratory,abel2022search}.
The results of these experiments are interpreted using models that assume that the UBDM is a virialized ensemble of non-interacting bosons described by the standard halo model (SHM) \cite{freese2013colloquium,pillepich2014distribution,evans2019refinement}.
These isotropic SHM UBDM models typically ignore any small-scale structure in the dark matter halo, beyond the stochastic fluctuations of the UBDM due to its finite coherence time \cite{foster2018revealing,centers2021stochastic,lisanti2021stochastic,gramolin2022spectral}.
Thus the sensors in these experiments are assumed to be quasi-continuously bathed in the UBDM field.

\begin{figure*}
\center
\includegraphics[scale=.7]{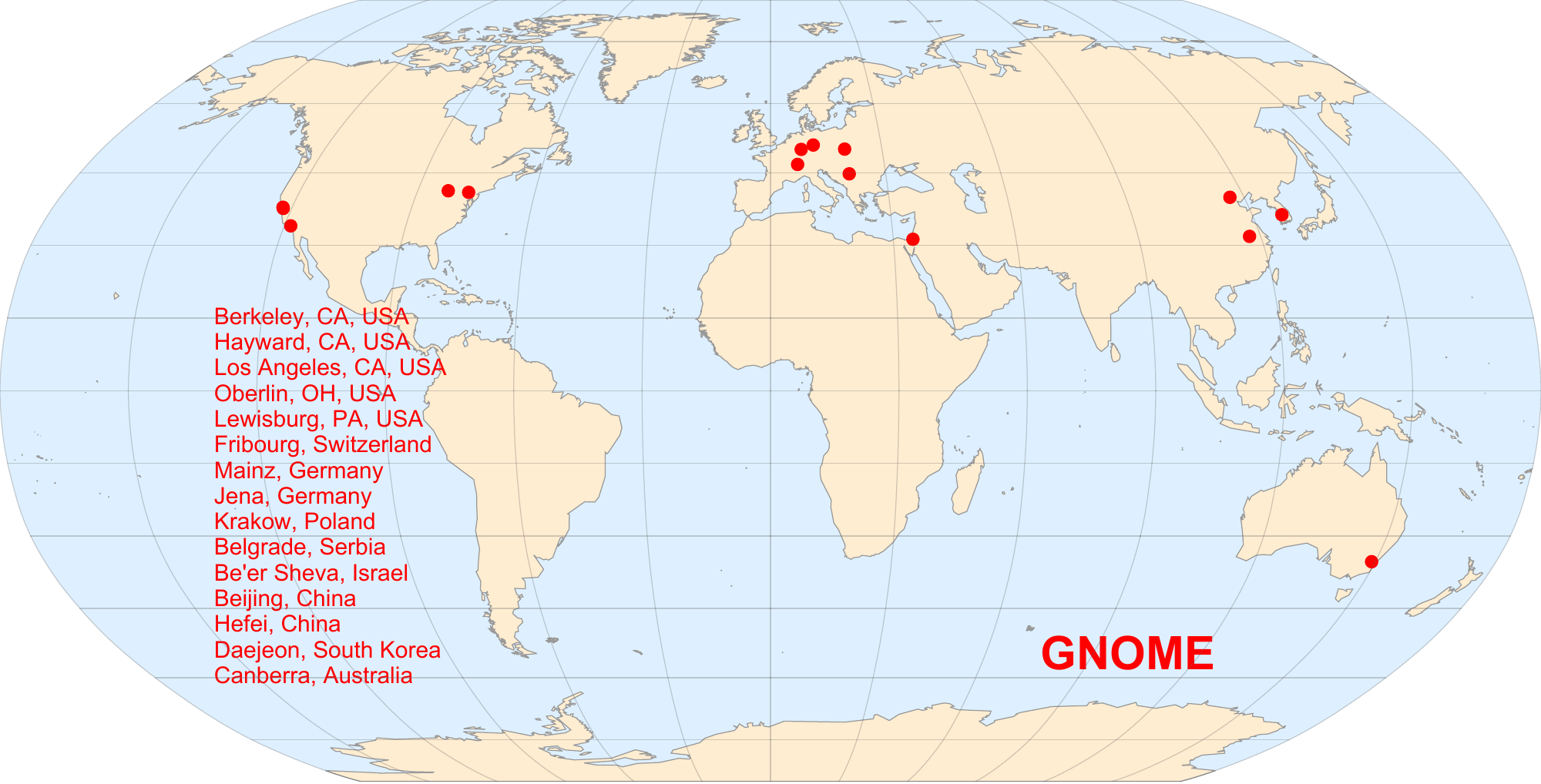}
\caption{Map and list of locations of GNOME stations. Note that the Fribourg station was moved to Jena in 2018.}
\label{Fig:GNOME-map}
\end{figure*}

The {\textbf{G}}lobal {\textbf{N}}etwork of {\textbf{O}}ptical {\textbf{M}}agnetometers for {\textbf{E}}xotic physics searches (GNOME) \cite{pustelny2013global,afach2018characterization,afach2021search} tests a different hypothesis.
Perhaps the energy density of the UBDM field is concentrated in large composite structures.
In this case, most of the time the Earth would be in a region of space where there is little or no dark matter \cite{Pos13}.
In this case, the Earth would only occasionally and briefly pass through dark matter, leading to rare and short-lived signals in dark matter detectors.
In principle, a single sensor could detect such transient events.
However, it would be exceedingly difficult to confidently distinguish a signal generated by an encounter with a composite UBDM structure from ``false positives''.
Such false positives can be induced by occasional abrupt changes of sensor operational conditions (such as those due to electronic noise spikes, laser mode hops, or vibrations).
The GNOME is a time-synchronized array of atomic magnetometers, widely distributed geographically (Fig.\,\ref{Fig:GNOME-map}).
The design of GNOME enables vetoing of false positive events, suppression of uncorrelated noise, and confident identification of transient signals attributable to exotic, beyond-the-Standard-Model physics.

%----------------------------------------------------------------
%----------------------------------------------------------------

In this paper, we review a variety of theoretical models describing sources of transient signals potentially detectable with GNOME.
We focus in particular on phenomenological descriptions of the temporal characteristics of the signals that would manifest in the GNOME sensors, which informs our data-analysis strategies.

%%%%%%%%%%%%%%%%%%%%%%%%%%%%%%%%%%%%%%%%%%%%%%%
%%%%%%%%%%%%%%%%%%%%%%%%%%%%%%%%%%%%%%%%%%%%%%%
\section{Interactions between atomic spins and ultralight bosonic fields}
\label{sec:spin-dependent-interactions}
%%%%%%%%%%%%%%%%%%%%%%%%%%%%%%%%%%%%%%%%%%%%%%%
%%%%%%%%%%%%%%%%%%%%%%%%%%%%%%%%%%%%%%%%%%%%%%%

The optical atomic magnetometers (OAMs) \cite{Bud13,budker2007optical} comprising GNOME are sensitive to interactions of atomic spins $\bs{F}$ with hypothetical exotic fields $\bs{\Upsilon}$, as well as magnetic fields $\bs{B}$, where $\bs{F}$ is the total atomic angular momentum.
For the exotic spin-dependent interactions considered here, the Hamiltonian has the form:
\begin{align}
    \mc{H}_\Upsilon = -\sum_{i=e,p,n}{g_{\Upsilon i} \frac{\bs{S}_i}{\left| S_i \right|} \cdot \bs{\Upsilon}} = -\sum_{i}{g_{\Upsilon i} \sigma_i \frac{\bs{F}}{\left| F \right|} \cdot \bs{\Upsilon}}\,,
    \label{eq:exotic-field-Hamiltonian}
\end{align}
where $g_{\Upsilon i}$ is the coupling constant characterizing the interaction of $\bs{\Upsilon}$ with the fermion spin [where the fermions considered include electrons ($i=e$), protons ($i=p$), and neutrons ($i=n$)], $\sigma_i$ is the fractional fermion spin polarization for a given atom (see the Supplemental Material in Ref.~\cite{afach2021search}), $\bs{F}$ is the total atomic angular momentum of the atomic state probed, and $\left| S_i \right| = 1/2$ and $\left| F \right|$ are the maximum spin projections.
The exotic-field Hamiltonian $\mc{H}_\Upsilon$ can be compared to the Zeeman Hamiltonian describing the interaction of atomic spins with a magnetic field:
\begin{align}
    \mc{H}_B = g_F \mu_B \bs{F} \cdot \bs{B}\,,
    \label{eq:Zeeman-Hamiltonian}
\end{align}
where $g_F$ is the Land\'{e} g-factor, $\mu_B$ is the Bohr magneton, and $\bs{B}$ is the external magnetic field experienced by the atom.
Comparing Eqs.\,\eqref{eq:exotic-field-Hamiltonian} and \eqref{eq:Zeeman-Hamiltonian}, it is evident that the physical manifestations of the ordinary magnetic field coupling and the coupling of exotic fields to spins are analogous.
Furthermore, by measuring OAM response to $\bs{B}$, the response to an exotic field $\bs{\Upsilon}$ can be inferred \cite{padniuk_response_2022}.
Therefore, we can consider $\bs{\Upsilon}$ to be a ``pseudo-magnetic'' field: a field that shifts Zeeman energy levels and generates torques on atomic spins, but does not couple proportionally to spins of different species according to their gyromagnetic ratios (i.e., comparing across different atomic species, $\sum_i g_{\Upsilon i} \sigma_i / \left| F \right|$ is not proportional to $g_F \mu_B$). Furthermore, unlike a magnetic field, it may be the case that $\bs{\Upsilon}$ has nonzero divergence (see, for example, Ref.\,\cite{grabowska2018detecting}).

In order to reduce environmental noise from ambient magnetic fields, the atomic vapor cells that contain the gases at the heart of GNOME's OAMs are placed inside multilayer magnetic shields composed of soft ferrimagnetic or ferromagnetic materials (such as mu-metal)  \cite{yashchuk2011magnetic}.
As noted in Ref.\,\cite{kimball2016magnetic}, if the exotic field $\bs{\Upsilon}$ interacts primarily with electron spins, there is an approximate cancellation of the effect of the field $\bs{\Upsilon}$ on electron spins within the magnetic shield.
This is due to the fact that it is the electron spins within shielding materials such as mu-metal that respond to the external magnetic field, so they similarly respond to an electron-coupled field $\bs{\Upsilon}$.
This response generates a corresponding magnetic field approximately cancelling the electron-spin-dependent energy shift within the shield \cite{kimball2016magnetic}.\footnote{Note that cancellation of the effects of an electron-spin-coupled $\bs{\Upsilon}$ is not exact within the ferromagnetic shielding due to imperfect shielding and nonzero nuclear magnetic moments. If the techniques of noble-gas-alkali-metal comagnetometry are employed (see Sec.~\ref{sec:Advanced}), electron-spin-coupled fields $\bs{\Upsilon}$ are measurable inside the shields because the nuclear spins of the noble gas respond to the $\bs{\Upsilon}$-induced magnetic field from the shielding material.}
Consequently, GNOME OAMs are primarily sensitive to exotic field couplings to nuclear spins.
At present, GNOME OAMs use alkali atoms such as rubidium (Rb) and cesium (Cs) whose nuclei have valence protons, and thus GNOME predominantly measures interactions of exotic fields with the proton spin $\bs{S}_p$ \cite{Kim15,stadnik2015nuclear}.

The principal hypothesis GNOME has sought to test is that dark matter is composed of ultralight spin-0 bosons known as axions or axion-like particles (ALPs).
Such exotic spin-0 bosons are ubiquitous features of the theoretical landscape beyond the Standard Model.
The axion originally emerged from a proposed solution to the strong-CP problem \cite{Pec77a,Pec77b,Wei77,Wil77}, the mystery of why nucleon EDMs and CP-violating nuclear electromagnetic moments are many orders of magnitude smaller than nominally predicted by quantum chromodynamics (QCD).
Since then, a variety of other beyond-the-Standard-Model theories have emerged predicting similar spin-0 bosons known as ALPs \cite{kimball2022search,Gra15,co2021predictions,Svr06,Arv10}.
Axions and ALPs are commonly thought to be ultralight (masses $m_a \ll 1\,{\rm eV}$).
They can be copiously produced in the early universe \cite{Abb83,Pre83,Din83,Fre10,Cha98,Ark09,arias2012wispy} and have all the requisite characteristics to be the dark matter \cite{graham2015experimental,graham2018stochastic,arias2012wispy,duffy2009axions}.

%For example, relaxions were proposed to solve the hierarchy problem \cite{Gra15}, the question of why the Higgs boson mass is so much lighter than the Planck mass (or, in other words, why the electroweak interaction so much stronger than gravity).
%Axions and ALPs also offer a mechanism to explain the asymmetry between matter and antimatter in the universe \cite{co2021predictions}.
%Furthermore, attempts to unify general relativity and quantum field theory, such as string theories, generically predict the existence of ALPs \cite{Svr06,Arv10}.
%The key point is that new spin-0 bosons are extremely well-motivated from a wide variety of theoretical perspectives.

%Axions and ALPs are characterized by a spontaneous symmetry breaking scale $f_a$ and an interaction scale $\Lambda$ (that is responsible for soft explicit symmetry breaking).
%Symmetry breaking scales of particular interest are the grand unified theory (GUT) scale at $\sim 10^{16}~{\rm GeV}$ and the Planck scale at $\sim 10^{19}~{\rm GeV}$.
%For the standard QCD axion the characteristic interaction scale is $\Lambda \sim 200~{\rm MeV}$, but for ALPs $\Lambda$ is essentially unconstrained \cite{kim2010axions}. These two scales determine the axion/ALP mass
%\begin{align}
%m_a c^2 = \frac{\Lambda^2}{f_a}~.
%\end{align}

The most commonly considered manifestation of a coupling between an ALP field $\varphi$ and atomic spins is given by the Lagrangian \cite{graham2013new}
\begin{align}
    \mc{L}_l = \frac{\prn{\hbar c}^{3/2}}{f_l} J^\mu \partial_\mu \varphi\,,
    \label{eq:linear-interaction-Lagrangian}
\end{align}
where $f_l$ is the characteristic energy scale associated with the spin ``portal'' between ALPs and fermions (the subscript $l$ denoting that the interaction is linear in the ALP field $\varphi$) and $J^\mu$ is the axial-vector current for fermions $\psi$,
\begin{align}
    J^\mu = \bar{\psi} \gamma^\mu\gamma_5 \psi\,,
    \label{eq:axial-vector-current}
\end{align}
where $\gamma^\mu$ and $\gamma_5$ are Dirac matrices.
The corresponding Hamiltonian $\mc{H}_l$ can be derived from the Euler-Lagrange equations (see, for example, Refs.\,\cite{kimball2022search,dailey2021quantum}):
\begin{align}
    \mc{H}_l \psi = -\frac{\prn{\hbar c}^{3/2}}{f_l} \gamma_0 \gamma^\mu\gamma_5 \prn{\partial_\mu \varphi} \psi~.
    \label{eq:linear-Hamiltonian-relativistic}
\end{align}
In the nonrelativistic limit, where the spacelike component of the derivative of $\varphi$ is much larger than the timelike component,
\begin{align}
    \mc{H}_{li} = -\frac{\prn{\hbar c}^{3/2}}{f_{li}} \frac{\bs{S}_i}{\left| S_i \right|} \cdot \bs{\nabla} \varphi~,
    \label{eq:linear-Hamiltonian}
\end{align}
where the subscript $i$ specifies the interaction with fermion $i=e,p,n$.
Comparing Eq.\,\eqref{eq:linear-Hamiltonian} to Eq.\,\eqref{eq:exotic-field-Hamiltonian}, we see that the coupling constant for ALPs in the above parameterization is given by $g_{\Upsilon i} = \prn{\hbar c}^{3/2}/f_{li}$ and the exotic pseudo-magnetic field is described by $\bs{\Upsilon}_l = \bs{\nabla} \varphi$.
Note that not only does $\mc{H}_l$ generate an interaction between spins and the spatial gradient of $\varphi$, but $\mc{H}_l$ also generates an interaction between spins that move with respect to the ALPs, since the momentum is related to the gradient operator via $\bs{p} = -i\hbar \bs{\nabla}$.
The latter interaction is often referred to as the ``axion wind'' \cite{Flambaum:2013Axion_Patras,graham2013new,stadnik2014axion,jackson2020overview}.

We also consider an alternative ``quadratic'' coupling between spins and the gradient of the intensity of the ALP field, $\bs{\nabla}\varphi^2$ \cite{Pos13}.
Whereas the QCD axion associated with the Peccei-Quinn solution to the strong-CP problem \cite{Pec77a,Pec77b} generally possesses the linear gradient interaction described by Eqs.\,\eqref{eq:linear-interaction-Lagrangian} and \eqref{eq:linear-Hamiltonian} \cite{graham2013new}, the quadratic gradient interaction can arise in effective field theories predicting ALPs not associated with the QCD sector \cite{Oli08}.
It is possible that under certain circumstances (see, for example, Refs.~\cite{damour1993general,damour1993tensor,damour1994string}), linear-in-$\varphi$ interactions may be suppressed, and the interaction of photons, electrons and nuclei with scalar fields starts at the quadratic order, $\varphi^2$. A very important consequence of such a modification is the relaxation of the most stringent astrophysics bounds compared to the linear case \cite{Oli08}, opening up a parameter space for the direct searches of $\varphi^2$ coupling to spins.
Also note that quadratic-in-$\varphi$ interactions are required for complex-valued $\varphi$, as is the case for the Q-ball scenario discussed in Sec.\,\ref{subsec:Q-balls}.
Furthermore, there are novel experimental signatures and modalities that can be employed to search for the quadratic-in-$\varphi$ interactions \cite{masia2022intensity}.

The Lagrangian describing the quadratic gradient coupling between an ALP field and atomic spins is given by \cite{Pos13}
\begin{align}
    \mc{L}_q = \frac{\hbar^2 c^2}{f_q^2} J^\mu \partial_\mu \varphi^2~.
    \label{eq:quadratic-interaction-Lagrangian}
\end{align}
The corresponding Hamiltonian is described by
\begin{align}
    \mc{H}_q \psi = -\frac{\hbar^2 c^2}{f_q^2} \gamma_0 \gamma^\mu\gamma_5 \prn{\partial_\mu \varphi^2} \psi\,,
    \label{eq:quadratic-Hamiltonian-relativistic}
\end{align}
and in the nonrelativistic limit the Hamiltonian describing the interaction of $\varphi$ with the spin of fermion $i$ is approximately
\begin{align}
    \mc{H}_{qi} = -\frac{\hbar^2 c^2}{f_{qi}^2} \frac{\bs{S}_i}{\left| S_i \right|} \cdot \bs{\nabla} \varphi^2~.
    \label{eq:quadratic-Hamiltonian}
\end{align}
Comparing Eq.\,\eqref{eq:quadratic-Hamiltonian} to Eq.\,\eqref{eq:exotic-field-Hamiltonian}, we see that the coupling constant for ALPs in the above parameterization is given by $g_{\Upsilon i} = \prn{\hbar c}^{2}/f_{qi}^2$ and the exotic pseudo-magnetic field is described by $\bs{\Upsilon}_q = \bs{\nabla} \varphi^2$.\footnote{Note that, as defined here, the units of $\bs{\Upsilon}_q = \bs{\nabla} \varphi^2$ differ from those of $\bs{\Upsilon}_l = \bs{\nabla} \varphi$. The correct units for the associated Hamiltonian, described by Eq.~\eqref{eq:exotic-field-Hamiltonian}, are obtained through the respective coupling constants $g_\Upsilon$ also having different units for the quadratic and linear ALP gradient interactions.}

It is of interest to note the discrete symmetry properties of the interactions described by Eqs.\,\eqref{eq:linear-Hamiltonian} and \eqref{eq:quadratic-Hamiltonian}.
Consider as a reference the discrete symmetry properties of the standard Zeeman interaction of Eq.\,\eqref{eq:Zeeman-Hamiltonian}.
The atomic angular momentum $\bs{F}$ is even under parity (P-even), since it is an axial- or pseudo-vector, and odd under time-reversal (T-odd), since angular momentum reverses its sign when time runs backward.
The magnetic field $\bs{B}$, generated by current flow, is also P-even and T-odd, and thus $\mc{H}_B$ is P- and T-even.
For the linear gradient interaction of Eq.\,\eqref{eq:linear-Hamiltonian}, the pseudoscalar ALP field $\varphi$ is P-odd and T-odd and the gradient $\bs{\nabla}$ is P-odd and T-even, and thus $\bs{\nabla} \varphi$ is a P-even, T-odd quantity, matching the discrete symmetry properties of the magnetic field.
Consequently, $\mc{H}_l \propto \bs{S}\cdot\bs{\nabla} \varphi$ is P- and T-even.
On the other hand, for the quadratic gradient interaction of Eq.\,\eqref{eq:quadratic-Hamiltonian}, $\bs{\nabla} \varphi^2$ is P-odd and T-even because of the extra factor of the ALP field $\varphi$, which means that the quantity $\bs{S}\cdot\bs{\nabla} \varphi^2$ is P- and T-odd.
Based on CPT invariance (where C is the charge conjugation symmetry), it follows that $\mc{H}_q$ describes a CP-violating interaction, and as a result could play a role in baryogenesis \cite{bernreuther2002cp}.

%%%%%%%%%%%%%%%%%%%%%%%%%%%%%%%%%%%%%%%%%%%%%%%
%%%%%%%%%%%%%%%%%%%%%%%%%%%%%%%%%%%%%%%%%%%%%%%
\section{GNOME overview}
%%%%%%%%%%%%%%%%%%%%%%%%%%%%%%%%%%%%%%%%%%%%%%%
%%%%%%%%%%%%%%%%%%%%%%%%%%%%%%%%%%%%%%%%%%%%%%%

The idea of the GNOME experiment is to carry out synchronous measurements of spin-dependent interactions using OAMs operating within magnetically-shielded environments in distant locations.
In this section we review the basic feature of the GNOME network, give an overview of the data collected so far, and describe ongoing improvements to the GNOME sensors that will significantly enhance the network sensitivity.

%%%%%%%%%%%%%%%%%%%%%%%%%%%%%%%%%%%%%%%%%%%%%%%
\subsection{GNOME magnetometers}
%%%%%%%%%%%%%%%%%%%%%%%%%%%%%%%%%%%%%%%%%%%%%%%

As mentioned above, OAMs utilize the interaction of atomic spins with external magnetic fields \cite{Bud13}. Typically, alkali metal vapors, contained in glass cells, are used for the measurements.
To prevent spin-depolarizing collisions with cell walls, which can limit the OAM sensitivity, either the walls are coated with special (e.g., paraffin) layer or the cells are filled with an additional inert gas (e.g., noble gas) to slow down diffusion.
In OAMs, the atoms are optically polarized, resulting in optical anisotropy of the medium.
As the spins of the polarized atoms precess due to a nonzero magnetic field (or, perhaps, due to an exotic field coupled to atomic spins), detection of the corresponding change in optical properties of the medium provides quantitative information about the field.
The signals of the GNOME magnetometers are recorded using a custom data acquisition (DAQ) system \cite{Wlod2014}, providing accurate timing from the Global Positioning System (GPS).

In the first incarnation of the GNOME, various OAMs employed different elements (Rb and Cs) and were based on different techniques; spin-exchange-relaxation-free (SERF) \cite{Kominis03}, M$_x$ \cite{Greoger06}, and nonlinear-magneto-optical-rotation (NMOR) \cite{Pustelny08} magnetometers were used for the measurements.
On the one hand, this diversity offers flexibility, opens a greater range of theoretical parameter space for exploration, and improves the robustness of the network.
On the other hand, it results in sensors having different sensitivities and bandwidths, which complicates data analysis.
In general, however, magnetometers used in the GNOME have an operational sensitivity better than 1\,pT/$\sqrt{\text{Hz}}$, corresponding to a sensitivity to Zeeman energy shifts below 10$^{-17}$\,eV/$\sqrt{\text{Hz}}$, and bandwidths up to 100\,Hz \cite{afach2018characterization}.

Although OAMs enable searches for non-magnetic spin couplings, the devices are highly sensitive to magnetic fields.
Therefore, despite shielding from the external environment, uncontrollable magnetic disturbances are a significant source of noise.
In order to reduce magnetic noise, the next generation of the GNOME experiment (Advanced GNOME) is using comagnetometers \cite{kornack_dynamics_2002,terrano2022comagnetometer}: sensors with limited sensitivity to magnetic fields that still maintain sensitivity to non-magnetic spin couplings.
Comagnetometers are briefly described in Sec.\,\ref{sec:Advanced}.

%----------------------------------------------------------------
\begin{figure*}
\center
\includegraphics[width=4.5in]{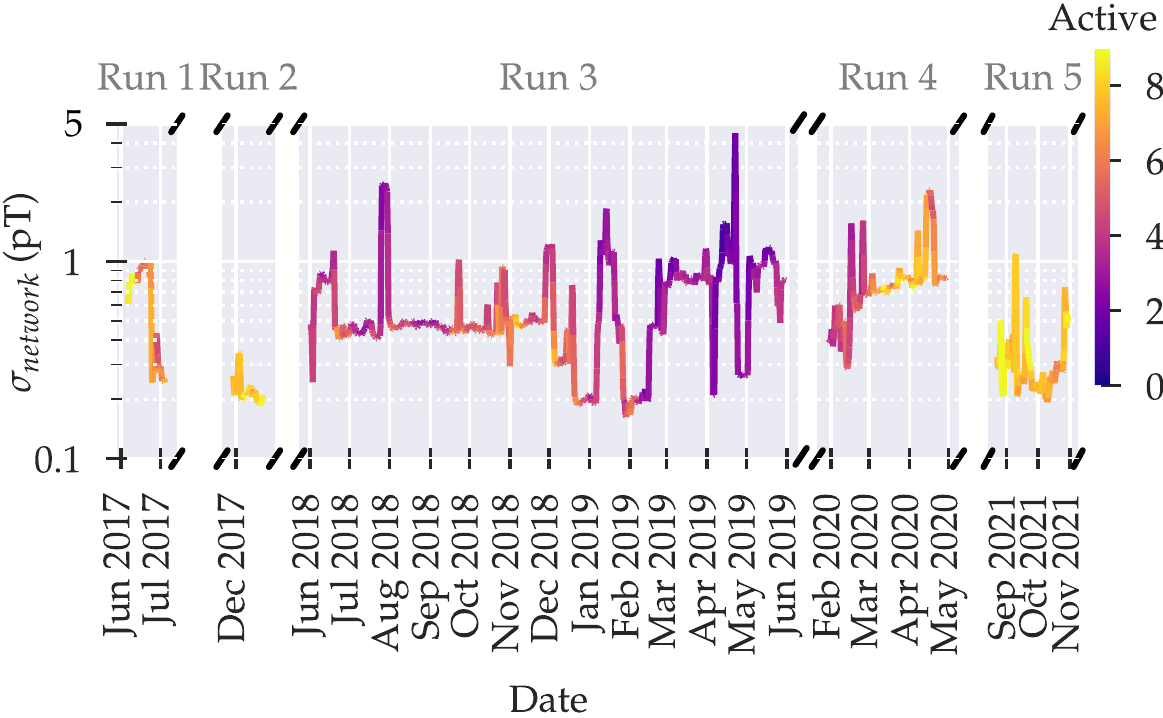}
\caption{One-day rolling average of the GNOME noise level according to Eq.\,\eqref{eq:sens} over the course of the first five Science Runs. The standard deviation at each magnetometer is calculated using the data for one second. Then this information is averaged for one hour. The color of the line indicates the number of active stations as indicated on the color map legend at the upper right.
}
\label{Fig:GNOME-sensitivity-over-time}
\end{figure*}
%----------------------------------------------------------------

%%%%%%%%%%%%%%%%%%%%%%%%%%%%%%%%%%%%%%%%%%%%%%%
\subsection{Monitoring of glitches}
%%%%%%%%%%%%%%%%%%%%%%%%%%%%%%%%%%%%%%%%%%%%%%%

While by its nature the GNOME can suppress uncorrelated noise and false positive events, additional measures are implemented in the network to further increase the data quality and trustworthiness of each station.
In order to veto signatures in the data which might have been produced by technical issues or changing experimental conditions (e.g., mechanical shocks, magnetic or electric pulses from neighboring technical devices), each GNOME station features a tailored automated system to continuously check for environmental perturbations.
The system is based on the Arduino microcontroller board MEGA 2560 additionally equipped with the Arduino 9 Axes motion shield (magnetometer, accelerometer and gyroscope, three axes each) in a separate ``sensor box''.
The sensor box is mounted on the optical table of the GNOME station near the magnetic shielding.
In this way, it can check for mechanical shocks or vibration of the optical setup.
In addition, the system features additional analog voltage inputs, which are used to monitor the operational status of the station (e.g., magnetometer signal amplitude, error signal of the magnetometer feedback and/or the laser lock, readings of temperature sensors, and signals of a photo diode monitoring laser or ambient light power).
For each GNOME station, depending on the specific setup, the system is set to monitor its critical parameters.

A dedicated Python-based software allows the system to display, save, and define ``sane'' ranges for all monitored parameters.
If one of the parameters falls out of the ``sane'' range, the system will output a signal to the DAQ system indicating the data recorded meanwhile should be rejected in data analysis.
The software writes a log file allowing one to trace back the event to the individual monitored parameter.
It is also possible to set alarms to notify station operators about irregularities over email and/or the Telegram app.
%

%%%%%%%%%%%%%%%%%%%%%%%%%%%%%%%%%%%%%%%%%%%%%%%
\subsection{Calibration pulses}
%%%%%%%%%%%%%%%%%%%%%%%%%%%%%%%%%%%%%%%%%%%%%%%

A possible concern with the continuous operation of the magnetometers over the course of several months is variation in the calibration and bandwidth of the detectors.
Such variations could result from drifts in laser power, laser frequency, or temperature of the vapor cells.
To monitor this, a series of oscillating magnetic fields are periodically applied to each magnetometer station via coils inside the magnetic shields.
The frequency of the applied magnetic field is stepped from 1 to 180\,Hz over the course of 9\,s using a programmable function generator.\footnote{The pulse sequence used in a recent Science Run 5 was 1\,Hz for 4\,s, 10\,Hz for 2\,s, 35\,Hz for 1\,s, 55\,Hz for 0.6\,s, 70\,Hz for 0.4\,s, 80\,Hz for 0.2\,s, 90\,Hz for 0.2\,s, 110\,Hz for 0.2\,s, 130\,Hz for 0.2\,s, 160\,Hz for 0.1\,s, and 180\,Hz for 0.1\,s.}
During the most recent experimental campaign (Science Run 5), the pulse sequence was applied hourly.
The response of the magnetometers at the different frequencies provides a convenient check on the operation of the magnetometers as well as a method for measuring the frequency response and bandwidth of the detectors.
The pulses also provide a test of the timing of the stations.
The pulses are triggered by the GPS pulse-per-second (pps) signal of the GNOME DAQ system or a time synchronized computer clock.
These tests indicate that the stations are synchronized at a level of better than the sample period, $T_\textrm{samp} = 1/512$\,s.\footnote{The GPS DAQ system provides timing with a precision better than 100\,ns.}

%%%%%%%%%%%%%%%%%%%%%%%%%%%%%%%%%%%%%%%%%%%%%%%
\subsection{The GNOME experiment so far}
%%%%%%%%%%%%%%%%%%%%%%%%%%%%%%%%%%%%%%%%%%%%%%%

To date, the GNOME collaboration has completed five ``Science Runs'' as well as a number of test and calibration runs.
According to Eqs.\,\eqref{eq:linear-Hamiltonian} and \eqref{eq:quadratic-Hamiltonian} and the surrounding discussion in Sec.\,\ref{sec:spin-dependent-interactions}, GNOME sensors seek to measure characteristic global patterns of pseudo-magnetic fields $\bs{\Upsilon}$.
Typically, GNOME magnetometers are sensitive to the projection of $\bs{\Upsilon}$ along a particular sensitive axis.
Thus the amplitudes of signals from exotic fields $\bs{\Upsilon}$ scale proportionally to $\hat{\bs{m}}\cdot\bs{\Upsilon}$, where $\hat{\bs{m}}$ is a unit vector pointing along the  sensitive axis of a magnetometer.
Since the direction of the field $\bs{\Upsilon}$ is essentially unknown, in order to assess the sensitivity of the GNOME, this directional sensitivity must be taken into account.
Additionally, the various parameters of individual sensors must be considered along with the relationship between the sensor response and the underlying physical theory (which must account for atomic and nuclear structure \cite{Kim15,stadnik2015nuclear}).
These issues are discussed in more detail in Refs.\,\cite{masia2020analysis,afach2021search,smiga2022assessing}.

A summary of the network performance for the five GNOME Science Runs is shown in Fig.\,\ref{Fig:GNOME-sensitivity-over-time}.
Since the GNOME data can be utilized in different ways to test different exotic physics hypotheses, discussed in Sec.\,\ref{sec:search-targets}, for simplicity we adopt for the summary a relatively simple, model-independent evaluation.
The plot in Fig.\,\ref{Fig:GNOME-sensitivity-over-time} shows the one-hour-average noise level as defined by
\begin{align}
\sigma\ts{network} \equiv \sqrt{ \frac{1}{ \sum_{j} \sigma_j^{-2}  } }\,,
\label{eq:sens}
\end{align}
where $\sigma_j$ is the variance of magnetometer $j$, calculated using the standard deviation for each second of data.
As can be seen from Fig.\,\ref{Fig:GNOME-sensitivity-over-time}, the GNOME experiment has accumulated over a year of data sensitive to pseudo-magnetic fields with equivalent magnitudes $\lesssim \textrm{pT}$ that can be searched for a variety of exotic physics signals.

%%%%%%%%%%%%%%%%%%%%%%%%%%%%%%%%%%%%%%%%%%%%%%%
\subsection{Advanced GNOME: noble gas comagnetometers}
\label{sec:Advanced}
%%%%%%%%%%%%%%%%%%%%%%%%%%%%%%%%%%%%%%%%%%%%%%%

%\DK{Lead writer: Misha/Arne/Manu}
%\MP{Part of the paragraf below might go to the GNOME magnetometers section of be removed according to the GNOME magnetometers section.}

The further development of the GNOME experiment focuses on the diversification and upgrade of the sensors implemented in the network.
There are three main directions for the improvement of sensors: enhancing their sensitivity, increasing their bandwidth, and expanding the types of couplings probed.
Although a number of various experimental techniques could be used for GNOME sensors (e.g., spin-based amplifiers \cite{jiang2022floquet}, noble gas masers \cite{Gle08,jiang2021floquet}, dual-species nuclear-spin comagnetometers \cite{Ven92,feng2022search}, alkali comagnetometers \cite{kimball2013dual,kimball2017constraints,*kimball2023GDMerratum,wang2020single}, liquid-state NMR comagnetometers operating in the zero-to-ultralow field (ZULF) regime \cite{ledbetter2012liquid,wu2018nuclear}, etc.), efforts are presently focused on developing self-compensating noble-gas-alkali-metal comagnetometers \cite{kornack_dynamics_2002,terrano2022comagnetometer} for implementation in the ``Advanced GNOME'' experiment \cite{padniuk_response_2022,klinger2022polarization}.

In addition to the coupling of exotic fields to proton spins, which was the only coupling probed at a competitive level by the first-generation GNOME \cite{afach2021search}, self-compensating noble-gas-alkali-metal comagnetometers can also probe both neutron and electron spin couplings. %The advanced GNOME sensor sensitivity to generic nuclear couplings is superior to the sensitivity of alkali metal magnetometers.
The ability to measure neutron spin couplings to $\bs{\Upsilon}$ comes from the fact that the noble gases (such as $^3$He) employed in these sensors have nuclei with valence neutrons \cite{Kim15}.\footnote{Note that $^3$He-alkali-metal comagnetometers retain sensitivity to proton couplings via a reasonably well-measured and understood proton-spin polarization in the $^3$He nucleus and the proton-spin polarization of the alkali metal nucleus \cite{Kim15}.}
However, the reason for the ability of these sensors to measure electron spin couplings is somewhat more subtle.
If we assume that $\bs{\Upsilon}$ couples primarily to electron spins, then the ferromagnetic (or ferrimagnetic) shielding responds to $\bs{\Upsilon}$ by creating an induced magnetic field that cancels the effect of $\bs{\Upsilon}$ on electrons within the shield where the vapor is located \cite{kimball2016magnetic}.
However, if we assume no coupling of $\bs{\Upsilon}$ to neutron spins, then neutrons will respond to the induced magnetic field from the shields and thus a detectable effect is generated. This further highlights the advantages of comagnetometry for exotic physics searches with the GNOME.
%\DKDel{This is mainly due to the high concentration of the hyperpolarized nuclear spins of the noble gas, which effectively gains the effect of the exotic spin couplings on the alkali metal spins, which are used for the readout of the comagnetometer.}
At sub-Hz frequencies, the sensitivity of the comagnetometer is also significantly improved by the suppression of the magnetic field response due to the self-compensation regime in which the comagnetometer is operated. %The self-compensation regime not only allows to overcome the problem of the magnetic field noise, but
%\DKDel{Furthermore, operation in the self-compensating regime prevents the interfering of the magnetic shields with the measurements of the electron spin couplings and therefore enables GNOME searches for exotic spin couplings to electron spins \cite{kimball2016magnetic}.}

\begin{figure*}[htb]
    \centering
    \includegraphics[width=0.49\textwidth]{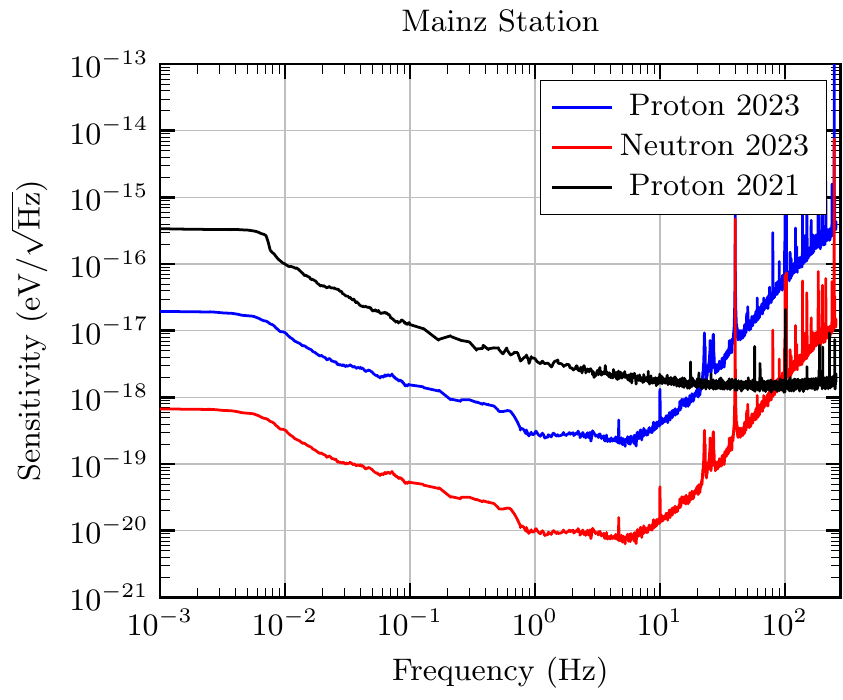}
    \includegraphics[width=0.49\textwidth]{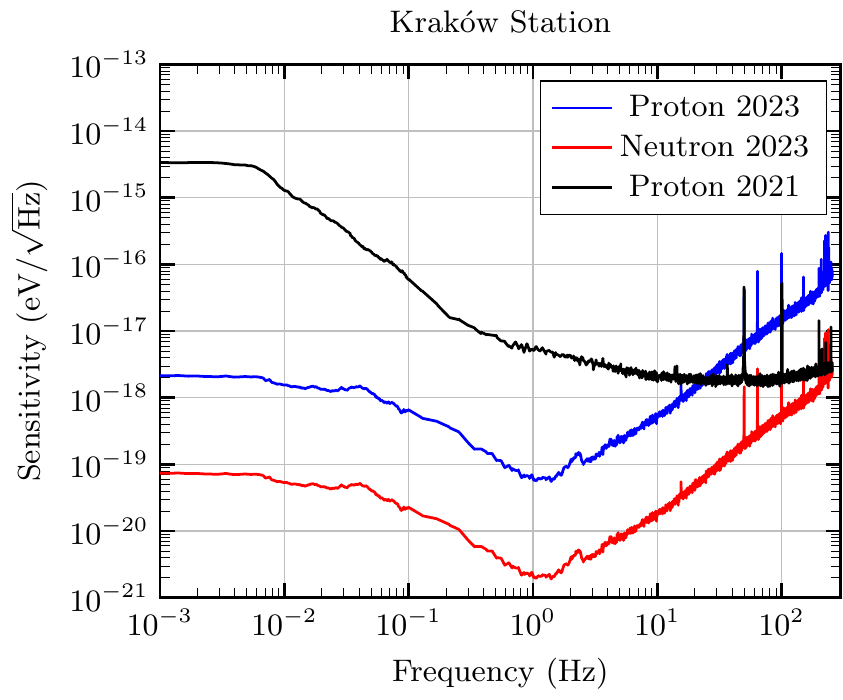}
    \caption{Sensitivities of Advanced GNOME stations in Mainz (left) and Krak\'ow (right) obtained during a Test Run in January 2023, compared with the typical sensitivity of the first-generation GNOME stations during Science Run 5 (see Fig.~\ref{Fig:GNOME-sensitivity-over-time}). \textbf{Red curves}: sensitivities of the comagnetometers to the neutron exotic spin couplings; \textbf{blue curves}: sensitivity of the comagnetometers to the proton spin couplings; \textbf{black curves}: magnetometer sensitivities to exotic proton couplings. The sensitivities are presented in terms of Zeeman-like energy splitting generated by the considered perturbation.  The presented data sets were undersampled for illustration purposes.}
    \label{fig:advancedGnome}
\end{figure*}

%The demonstrated sensitivity of the advanced GNOME sensors reaches the level of
%Test runs give a sensitivity of 10$^{-21}$ eV/$\sqrt{\mathrm{Hz}}$ (at 1 Hz) in case of the coupling to the neutron spins, see Fig.\,\ref{fig:advancedGnome} and about 10$^{-19}$ eV/$\sqrt{\mathrm{Hz}}$ (at 1 Hz) in case of the proton coupling.
Initial tests of Advanced GNOME sensors %described in Ref.\,\cite{klinger2022polarization}
demonstrate a sensitivity at the level of 10$^{-21}$\,eV/$\sqrt{\mathrm{Hz}}$ (at 1\,Hz) for exotic fields coupling to neutron spins,
and about 10$^{-19}$\,eV/$\sqrt{\mathrm{Hz}}$ (at 1\,Hz) for the proton spin coupling (surpassing that of GNOME magnetometers by a factor of 100), see Fig.\,\ref{fig:advancedGnome}.
% \DK{I still think it would be SO cool to have that figure showing the comagnetometer sensitivity here, but it is up to you all if you want to publish that here.}
%The difference in the sensitivity to proton and neutron couplings arises from the fact that the nuclear spin of the noble gas is mostly determined by the nuclear spin content of the noble gases.
For the electron spin couplings, the expected sensitivity is comparable to the sensitivity of the first-generation GNOME magnetometers to the proton exotic spin couplings.
The noble-gas-alkali-metal comagnetometers have an optimal sensitivity to the nuclear spin couplings for frequencies below a few hertz.
However, enhanced sensitivity to exotic fields (as compared to the first-generation of GNOME magnetometers) is expected over the whole bandwidth of the ordinary magnetometers, even at frequencies for which the comagnetometer performance is sub-optimal.

The Advanced GNOME sensors also bring another significant qualitative advantage for exotic physics searches.
The difference in the response to the magnetic and non-magnetic spin couplings can be used to discriminate events of magnetic origin from those driven by non-magnetic spin couplings, just from a single sensor readout \cite{padniuk_response_2022}.
Although ultimate verification of a global exotic physics event will rely on the correlation between signals observed in multiple GNOME stations, more effective discrimination between magnetic and non-magnetic signals will enable efficient suppression of the ``false positive'' rate and therefore improve the overall sensitivity of the network.

%The GNOME network can be improved by implementing a new type of sensors that is less prone to magnetic-field noise. This can be achieved by implementation a noble-gas-alkali-metal comagnetometers. These sensors allow a sensitivity gain to exotic spin couplings of protons and neutrons of three to five orders of magnitude. Moreover, in certain conditions comagnetometers allow to distinguish between magnetic and nonmagnetic spin perturbations. This is a significant advantage of the system compared to the conventionally used OPMs, as it allows to discriminate nonmagnetic couplings events.

%%%%%%%%%%%%%%%%%%%%%%%%%%%%%%%%%%%%%%%%%%%%%%%
%%%%%%%%%%%%%%%%%%%%%%%%%%%%%%%%%%%%%%%%%%%%%%%
\section{Search targets and their signals}
\label{sec:search-targets}
%%%%%%%%%%%%%%%%%%%%%%%%%%%%%%%%%%%%%%%%%%%%%%%
%%%%%%%%%%%%%%%%%%%%%%%%%%%%%%%%%%%%%%%%%%%%%%%

By analyzing correlations between signals from the geographically separated magnetometers and comagnetometers comprising GNOME, it is possible to probe a wide variety of exotic beyond-the-Standard-Model hypotheses.
In this section we survey some exotic physics scenarios that can be searched for using GNOME data and highlight examples of their particular temporal signatures.

%%%%%%%%%%%%%%%%%%%%%%%%%%%%%%%%%%%%%%%%%%%%%%%
%%%%%%%%%%%%%%%%%%%%%%%%%%%%%%%%%%%%%%%%%%%%%%%
\subsection{Axion domain walls}
\label{subsec:domain-walls}
%%%%%%%%%%%%%%%%%%%%%%%%%%%%%%%%%%%%%%%%%%%%%%%
%%%%%%%%%%%%%%%%%%%%%%%%%%%%%%%%%%%%%%%%%%%%%%%

%----------------------------------------------------------------
\begin{figure}
\center
\includegraphics[width=3.25in]{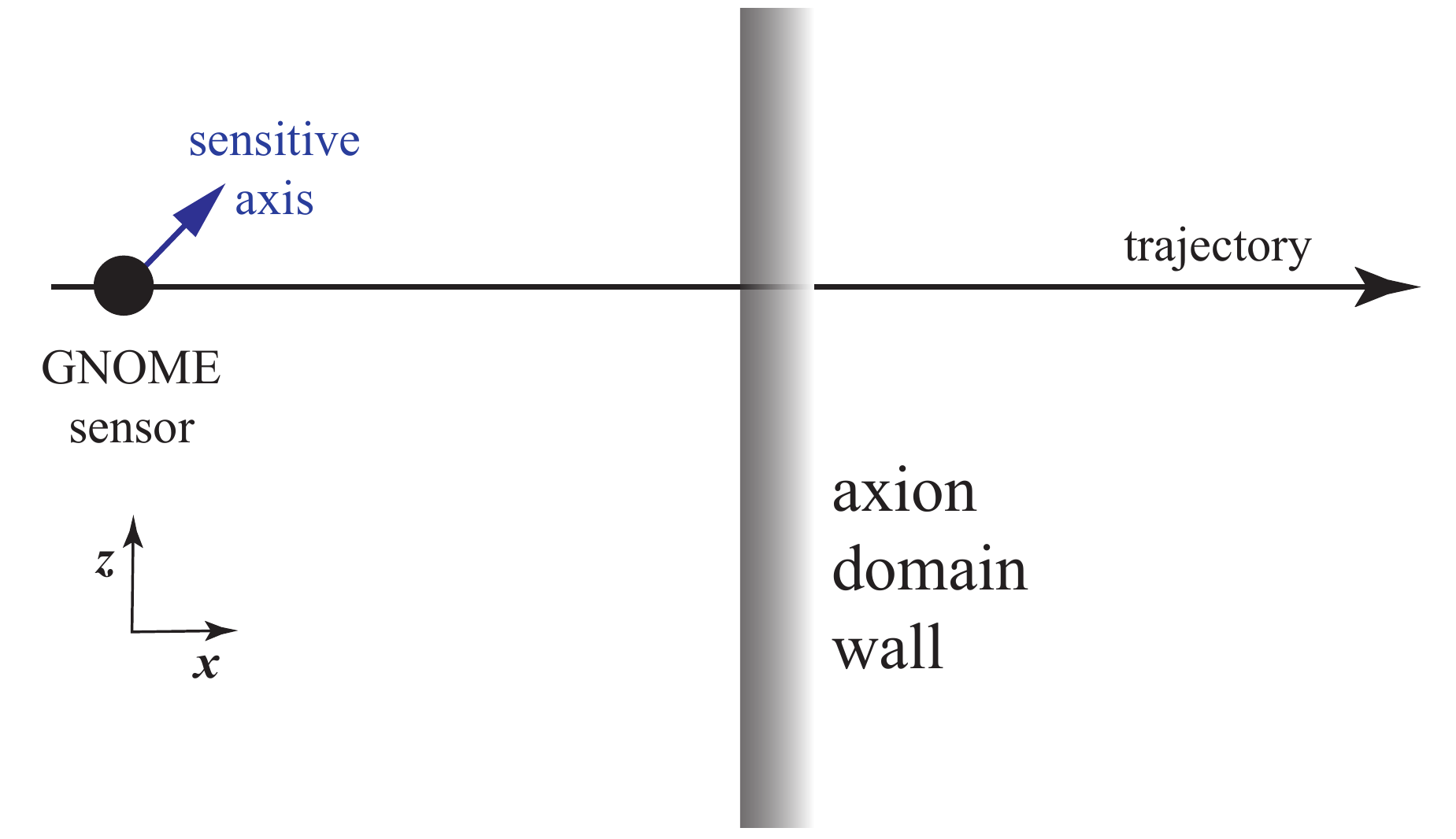}
\includegraphics[width=2.9in]{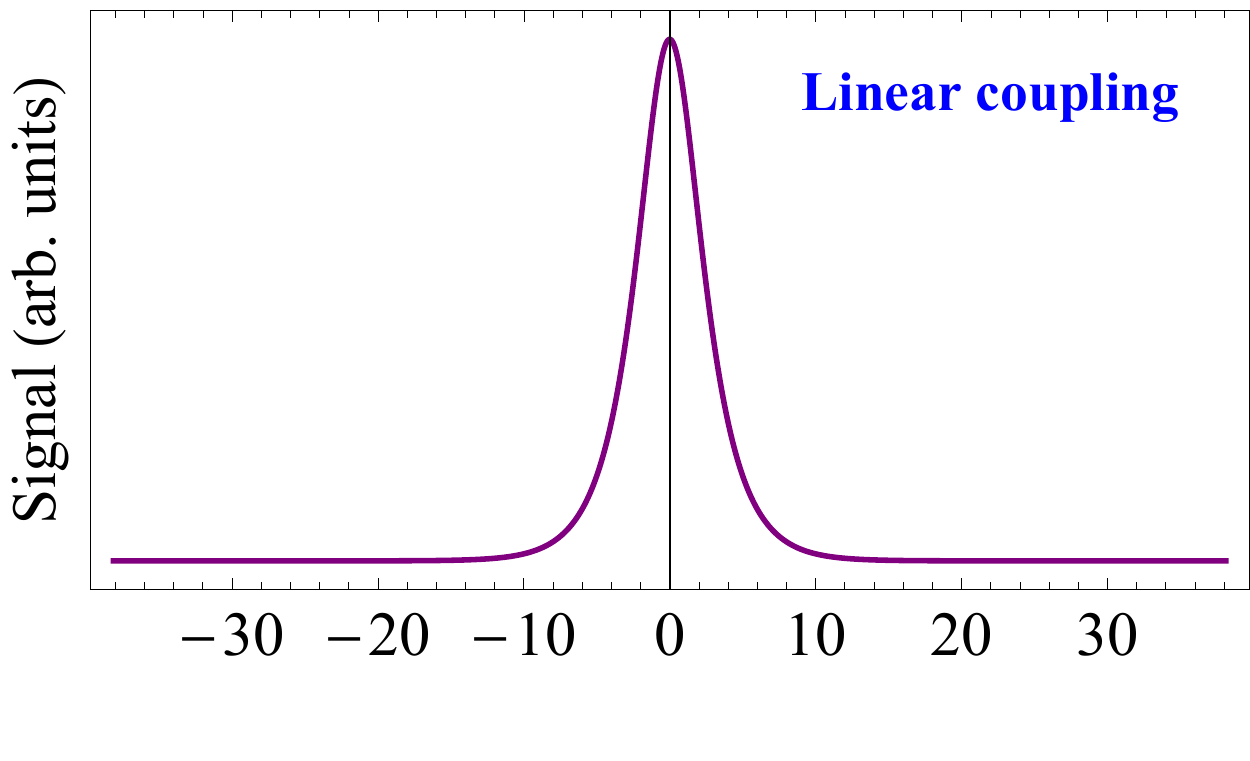}
\includegraphics[width=2.9in]{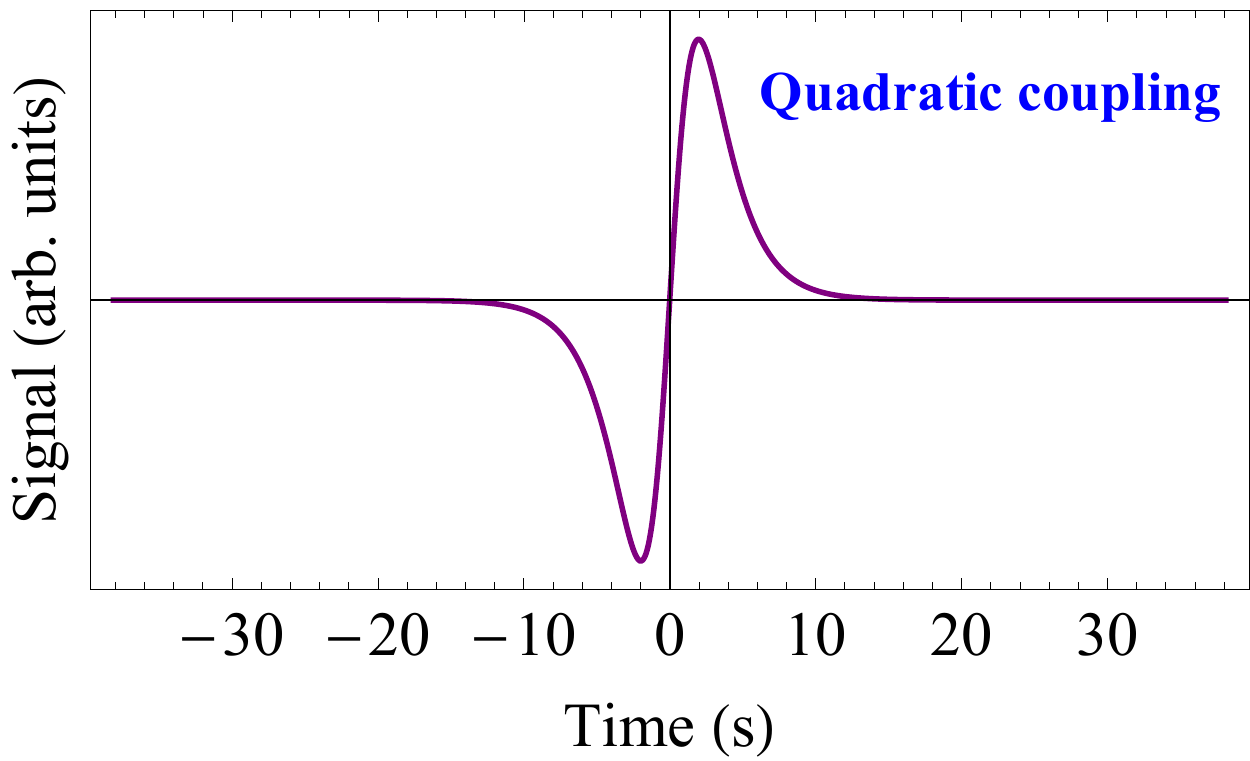}
\caption{Upper diagram: schematic depiction of a GNOME sensor passing through an ALP domain wall (thickness $\approx R_E/10$), indicating with the long black arrow the trajectory of the sensor along $\bs{v}$ (parallel to $\bs{k}$, along $\hat{\bs{x}}$, $v \approx 10^{-3}c$) and the sensitive axis of a GNOME sensor (blue arrow, tilted at $45^\circ$ to $\hat{\bs{x}}$). The plots below show example signals due to the linear spin coupling (middle plot) and quadratic spin coupling (lower plot) from an ALP-domain-wall encounter corresponding to the schematic at top.}
\label{Fig:domain-wall-signal}
\end{figure}
%----------------------------------------------------------------

%\DK{Lead writer: Daniel/Hector}

The first proposed search targets for GNOME were axion or ALP domain walls \cite{Pos13,pustelny2013global}.
Domain walls are topological defects that form between regions of space in which the ALP field possesses different, but energy-degenerate, vacuum states~\cite{sikivie1982axions,kawasaki2015axion}.
Such a scenario arises due to the non-trivial vacuum topology that ALP fields typically possess: there can be multiple local field energy minima (and corresponding vacuum states) that, in the abstract space describing the field, are not ``simply connected'' in a topological sense \cite{vilenkin1982cosmic}.
When spontaneous symmetry breaking occurs in the early universe, different regions of space (domains) acquire different vacuum states.
The domain walls are the field configurations at the boundaries between these domains.
In order for the ALP field to transition from one energy minimum to another across these boundaries, the ALP field necessarily acquires values above those corresponding to the minimum energy.
These non-vaccuum states are associated with considerable potential energy, and the change of the field over space means that the ALP field has a nonzero gradient.
%\DKDel{Since such value differs from the vacuum there is gradient of ALP field, and a tension is produced in an energy storage within the domain wall.}

%%%%%%%%%%%%%%%%%%%%%%%%%%%%%%%%%%%%%%%%%%%%%%%
\subsubsection{Theoretical description}
%%%%%%%%%%%%%%%%%%%%%%%%%%%%%%%%%%%%%%%%%%%%%%%

ALP domain walls are macroscopic field configurations that could be stable and long lived, and continue to exist today \cite{kawasaki2015axion}. They can be of astrophysical extent, potentially much larger than the size of Earth.
If the ALP field primarily manifests in the form of domain walls, the associated energy density would be concentrated into compact spatial regions.
While in most theoretical models such domain walls are unstable \cite{sikivie1982axions,nagasawa1994collapse}, in some other models \cite{Pos13,Der14,kawasaki2015axion} domain walls can compose a significant fraction of dark matter. These objects can be considered virialized in the galaxy according to the standard halo model (SHM). Based on this assumption, the rate and duration of the ALP domain wall encounters can be estimated. A region of theoretically plausible parameter space is expected to have an average encounter rate with Earth of a year or less \cite{Pos13,Der14}. This opens the possibility to directly search for such encounters. Atomic magnetometers and comagnetometers are sensitive to domain walls, since the ALP-field gradient can interact with the spin of electrons, neutrons and protons [see Eqs.\,\eqref{eq:linear-Hamiltonian} and \eqref{eq:quadratic-Hamiltonian}].

\subsubsection{Signal model}
%%%%%%%%%%%%%%%%%%%%%%%%%%%%%%%%%%%%%%%%%%%%%%%

Consider an ALP domain wall in the $yz$-plane ($x=0$) separating two degenerate ALP vacuum states as pictured in the upper diagram in Fig.\,\ref{Fig:domain-wall-signal}.
The solution of the field equations for $\varphi$ yields \cite{Zeldovich:1974uw,Pos13,afach2021search}
\begin{align}
    \varphi(x) = \varphi_0 \arcsin \sbrk{ \tanh\prn{ x/\lambda_c } }\,,
\end{align}
where $\varphi_0$ is a constant proportional to the spontaneous symmetry-breaking scale associated with the ALP and $\lambda_c = \hbar/\prn{m_a c}$ is the Compton wavelength of the ALP of mass $m_a$.
Atomic spins can interact with the ALP field through the linear and quadratic gradient interactions of Eqs.\,\eqref{eq:linear-Hamiltonian} and \eqref{eq:quadratic-Hamiltonian}, which can be described by the pseudo-magnetic fields
\begin{align}
    \bs{\Upsilon}_l = \bs{\nabla}\varphi(x)  = \frac{\varphi_0}{\lambda_c} \sech\prn{ x/\lambda_c } \,\hat{\bs{x}}
\end{align}
and
\begin{align}
    \bs{\Upsilon}_q = \bs{\nabla}\varphi^2(x) = \frac{2\varphi_0^2}{\lambda_c} \arcsin \sbrk{ \tanh\prn{ x/\lambda_c } } \sech\prn{ x/\lambda_c } \, \hat{\bs{x}}\,,
\label{Eq:quad_field}
\end{align}
respectively.
%{\color{blue} Shouldn't pseudo-magnetic fields include the coupling constants $f_l$ and $f_q^2$ from Eqs.\,\eqref{eq:linear-Hamiltonian} and \eqref{eq:quadratic-Hamiltonian}?
%Also, at the start of Sec.~\ref{sec:spin-dependent-interactions}, $\Upsilon$ is used to denote the exotic field itself, whereas here we refer to it as a pseudomagnetic field. }
%\DK{We have chosen by our definitions to use $\Upsilon$ as the ``exotic pseudo-magnetic field'' as opposed to the underlying pseudoscalar field $\varphi$. We also chose to not include the coupling constants in the field but rather treat them as ``pseudo-gyromagnetic ratio'' terms. I have attempted to make sure the language is consistent throughout and explained when first introduced.}
%{\color{blue}YS: Thanks for the explanation. It still looks like we use the terms ``exotic field'' and ``pseudo-magnetic field'' interchangeably for $\mathbf{\Upsilon}$ throughout the text. Hopefully this won't cause confusion for readers.}
The signal measured by a GNOME sensor is $\propto \hat{\bs{m}} \cdot \bs{\Upsilon}$, where $\hat{\bs{m}}$ is directed along the sensitive axis of the magnetometer.
The expected time-dependent lineshapes for the linear and quadratic couplings are shown in the lower plots of Fig.~\ref{Fig:domain-wall-signal}.
While all the GNOME sensors experience a common transient pseudo-magnetic field $\bs{\Upsilon}$, because of the varying directions of their sensitive axes $\hat{\bs{m}}$ (as well as differences in the magnetometers employed), the signals in different sensors would have varying amplitudes and signs.
For a given relative velocity between the ALP domain wall and Earth, there is a characteristic signal timing and amplitude pattern that can be used to distinguish true domain-wall-crossing events from spurious noise \cite{masia2020analysis,kim2022machine,afach2021search}.
Once an event is detected, one can relate the signal properties (width and the amplitude of the signal, as well as the inferred expected distribution of domain walls) to ALP parameters, including the mass of the ALP and the interaction and symmetry-breaking scales \cite{afach2021search}.

The GNOME collaboration has developed data analysis algorithms to search for ALP domain walls \cite{masia2020analysis,kim2022machine}, and has completed a full analysis of the data from Science Run 2 carried out in 2017 \cite{afach2021search}.
Our searches did not find any statistically significant signals above background that could point to the existence of ALP domain walls.
Consequently, our results can be interpreted as constraints on the properties of ALP domain walls.
The excluded (model-dependent) ALP parameter space covers masses in the range $\sim 10^{-15} - 10^{-7}$ eV \cite{afach2021search}.
A new analysis procedure is currently being developed to improve on the sensitivity to narrower domain wall widths with respect to our previous work \cite{afach2021search}. The new analysis is based on a preselection of candidate signals groups occurring within a given time window. Then each group is tested for consistency with the domain wall crossing model. This allows an efficient scrutiny of the data compared to the previous method. In Ref.~\cite{afach2021search} all possible domain wall configurations were scanned and their agreement with the data were evaluated. The new procedure results in a comparatively less computer-intensive routine.

%%%%%%%%%%%%%%%%%%%%%%%%%%%%%%%%%%%%%%%%%%%%%%%
%%%%%%%%%%%%%%%%%%%%%%%%%%%%%%%%%%%%%%%%%%%%%%%
\subsection{Axion stars}
\label{subsec:axion-stars}
%%%%%%%%%%%%%%%%%%%%%%%%%%%%%%%%%%%%%%%%%%%%%%%
%%%%%%%%%%%%%%%%%%%%%%%%%%%%%%%%%%%%%%%%%%%%%%%

%\DK{Lead writer: Jason/Derek}

Instead of the ALP dark matter being primarily in the form of topological defects such as domain walls as discussed above in Sec.~\ref{subsec:domain-walls}, it may be the case that inhomogeneities in the dark matter distribution provide seeds that enable ALP self-interactions or gravity to attract together a large local density of ALPs, thereby forming a spherical bound-state (see Refs.\,\cite{braaten2019colloquium,eby2019global,eby2021global,eby2018approximation} and references therein).
Such spherical bound states are referred to as axion or ALP stars, and they may constitute a significant fraction of the dark matter density \cite{braaten2019colloquium}.
Under certain conditions, ALP stars are stable under perturbations and radiative decay \cite{chavanis2011massAnalytic,chavanis2011massNumerical,eby2016lifetime,eby2018decay}, and can be efficiently formed in the early Universe \cite{schive2014understanding,levkov2018gravitational,eggemeier2019formation,kirkpatrick2020relaxation}.
The extensive theoretical studies of ALP stars to determine their evolution, stability, and characteristic radii and masses for different models of the ALP self-interaction establish that ALP stars are, indeed, a plausible dark matter scenario \cite{braaten2019colloquium,visinelli2018dilute,zhang2019axion}.
As described in Ref.\,\cite{Kim18AxionStars}, ALP stars are a compelling search target for the GNOME, since there is a range of masses and radii for ALP stars (not ruled out by existing empirical observations) for which the terrestrial encounter rate can be sufficiently high (at least once per year) that a detection would be feasible.
ALP stars can interact with atomic spins via the linear or quadratic gradient interactions [Eqs.\,\eqref{eq:linear-Hamiltonian} and \eqref{eq:quadratic-Hamiltonian}].
Thus if an ALP star passed through Earth, it would cause oscillatory pseudo-magnetic field pulses in GNOME sensors that, in principle, could be within the sensitivity range of GNOME \cite{Kim18AxionStars}.

%%%%%%%%%%%%%%%%%%%%%%%%%%%%%%%%%%%%%%%%%%%%%%%
\subsubsection{Theoretical description}
%%%%%%%%%%%%%%%%%%%%%%%%%%%%%%%%%%%%%%%%%%%%%%%

%----------------------------------------------------------------
\begin{figure}
\center
\includegraphics[width=3.35in]{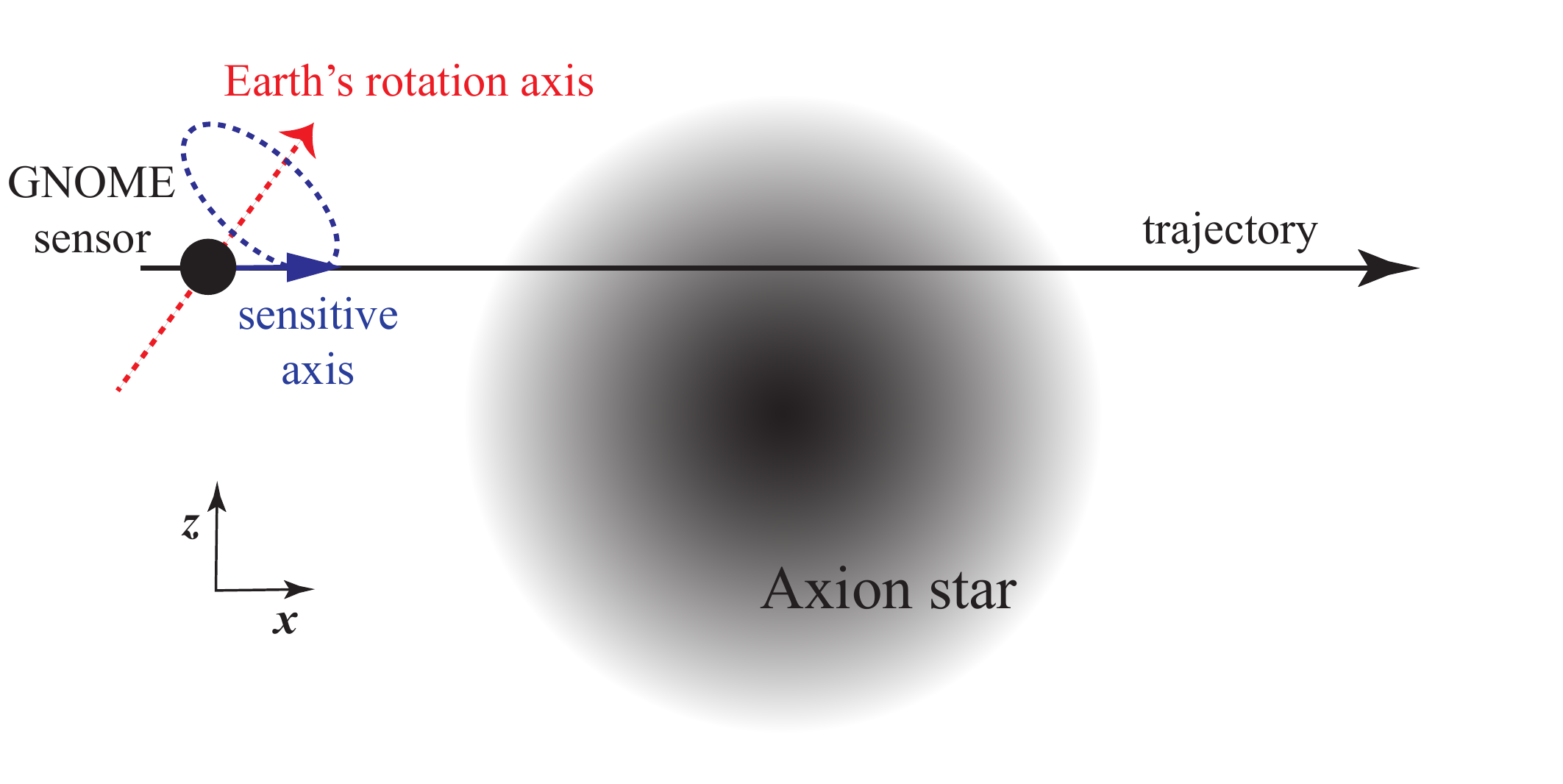}
\includegraphics[width=3.35in]{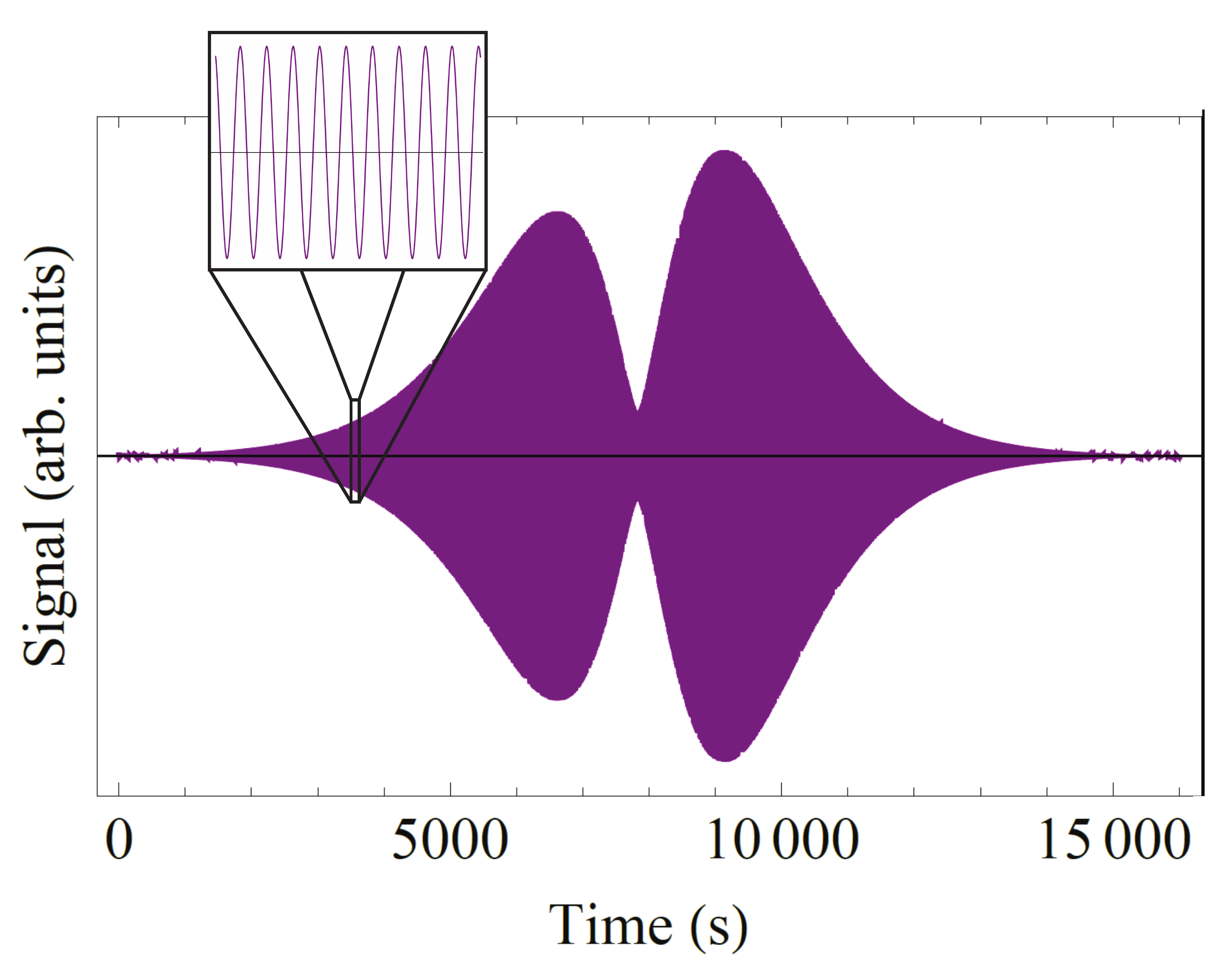}
\caption{Upper diagram: schematic depiction of a GNOME sensor passing through an ALP star, indicating with the long black arrow the trajectory of the sensor along $\bs{v}$ (parallel to $\bs{k}$, along $\hat{\bs{x}}$), the initial direction of the sensitive axis of the GNOME sensor (blue arrow, along $\hat{\bs{x}}$), and the axis of Earth's rotation (dashed red arrow, $45^\circ$ to the $x$-axis). Lower plot: example signal due to the linear spin coupling [Eq.\,\eqref{eq:ALPstarPseudoMagneticFieldLinear}] from an ALP star passage corresponding to the above schematic. For this example, the ALP star radius $R \approx 200 R_E$, $v = 10^{-3}c$, and at $t=0$ the sensor is located at $\prn{x_0,y_0,z_0} = \prn{2R,0,R/2}$, where the origin is defined to be the center of the ALP star. The ALP field oscillation frequency is chosen to be $\omega = 2\pi\times 10\,{\rm Hz}$, and we assume $\omega \approx \omega_c$, neglecting the binding energy. The ALP-star signal in the plot is based on the exponential model, Eq.\,\eqref{eq:ALPstar-exponential}. The inset plot shows an $\approx 1~{\rm s}$ long segment of the simulated signal.}
\label{Fig:axion-star-signal}
\end{figure}
%----------------------------------------------------------------

Axion or ALP stars held together via self-interactions are called {\emph{oscillons}} \cite{gleiser1994pseudostable} or {\emph{axitons}} \cite{Kol93}.
There have been a number of theoretical studies of such composite systems, and a few general features are observed.
Models of oscillons and axitons demonstrate that they can persist in a stable or quasi-stable regime \cite{honda2002fine,gleiser2004d,Mukaida:2016hwd}, however if the ALP density is too large, they tend to radiate ALP waves and lose mass \cite{braaten2019colloquium}.
Sufficiently dilute ALP stars can be stable over long time scales \cite{honda2002fine,gleiser2004d,mukaida2017longevity}.
The ALP fields bound in the form of oscillons have a characteristic field oscillation frequency $\omega$ which is somewhat smaller than the ALP Compton frequency $\omega_c$.
This stems from the fact that, while free ALPs oscillate at the Compton frequency $\omega_c$, because the energy of the ALPs in the star is reduced by a binding energy $\epsilon_b$, the oscillation frequency of ALPs in the oscillon is given by
\begin{align}
\omega = \omega_c - \epsilon_b/\hbar~.
\end{align}
This is a key difference between the ALP domain walls considered in Sec.~\ref{subsec:domain-walls} and the ALP stars considered here: the ALP field of the domain wall does not oscillate, being a topological defect related to the ALP field potential energy, whereas the ALP field of the star does oscillate, since most of the field energy is in the form of kinetic energy.
The mass $M$ of an oscillon is $M \approx N\prn{m_a - \epsilon_b/c^2}$, where $N$ is the number of ALPs in the star.

Alternatively, axion stars could be held together by gravity \cite{Kaup:1968zz,Ruf69,Sch03,Bar11,chavanis2011massAnalytic,chavanis2011massNumerical,eby2016lifetime,eby2018decay}.
However, in this case it may be expected that gravitational tidal forces due to the Sun, Earth, and other bodies in the solar system could cause distortions and disruptions of the axion star \cite{Tin16}, complicating the details of the shape profile in ways that are difficult to predict.
Thus, for simplicity, here we focus on the oscillon or axiton models where self-interactions stronger than gravity dominate.

%%%%%%%%%%%%%%%%%%%%%%%%%%%%%%%%%%%%%%%%%%%%%%%
\subsubsection{Signal model}
%%%%%%%%%%%%%%%%%%%%%%%%%%%%%%%%%%%%%%%%%%%%%%%

A number of different approximate analytic wavefunctions for ALP stars have been studied in the literature (see Ref.\,\cite{eby2018approximation} for a review).
Two of the more widely used radial wavefunctions $\Psi(r)$, where $r$ is the radial distance from the center of the ALP star, are the Gaussian approximation,
\begin{align}
    \Psi(r) = \sqrt{ \frac{N \kappa^3}{\pi^{3/2}R^3}} e^{-\kappa^2 r^2/(2 R^2)}\,,
    \label{eq:ALPstar-Gaussian}
\end{align}
and the exponential approximation,
\begin{align}
    \Psi(r) = \sqrt{ \frac{N \kappa^3}{\pi R^3}} e^{-\kappa r/R}\,,
    \label{eq:ALPstar-exponential}
\end{align}
where $R$ is the characteristic radius of the ALP star enclosing 99\% of its mass and $\kappa$ is a numerical parameter ($\kappa = 2.8$ for the Gaussian wavefunction and $\kappa = 4.2$ for the exponential wavefunction \cite{eby2018approximation}).
In Ref.\,\cite{eby2018approximation} it is argued that, in fact, a more accurate description of an ALP star in many relevant scenarios is given by a ``linear-plus-exponential'' wave function:
\begin{align}
    \Psi(r) = \sqrt{ \frac{N \kappa^3}{7 \pi R^3}} \prn{1 + \frac{\kappa r}{R}} e^{-\kappa r/R}~.
    \label{eq:ALPstar-linear-plus-exponential}
\end{align}
In any case, the ALP field as observed from the reference frame of a GNOME sensor is given by
\begin{align}
    \varphi(\bs{r},t) = \Psi(r)\cos\prn{ \bs{k}\cdot\bs{r} - \omega t + \theta }\, ,
\label{eq:ALPstarField}
\end{align}
where $\bs{k}$ is the ALP wave vector, $\bs{r}$ is the distance from the center of the ALP star to the sensor, and $\theta$ is a random phase.
Since the ALP field in the star is essentially a Bose condensate with long coherence time \cite{braaten2019colloquium}, we can treat the phase $\theta$ as constant throughout the duration of the passage of the GNOME sensor through the ALP star, and for simplicity in the following discussion we set $\theta = 0$.

The expected signal in a GNOME sensor can be calculated based on the exotic pseudo-magnetic fields $\bs{\Upsilon}_l = \bs{\nabla} \varphi$ and $\bs{\Upsilon}_q = \bs{\nabla} \varphi^2$ obtained from the expression \eqref{eq:ALPstarField}, where (as above) the subscripts denote the linear or quadratic fields:
\begin{align}
    \bs{\Upsilon}_l & = \prn{ \bs{\nabla} \Psi } \cos\prn{ \bs{k}\cdot\bs{r} - \omega t} - \bs{k} \Psi(r) \sin\prn{ \bs{k}\cdot\bs{r} - \omega t}\,, \nonumber \\
    & = \pdbyd{\Psi}{r}\cos\prn{ \bs{k}\cdot\bs{r} - \omega t} \hat{\bs{r}} - \frac{m_a v}{\hbar} \Psi(r) \sin\prn{ \bs{k}\cdot\bs{r} - \omega t} \hat{\bs{k}}\,,
\label{eq:ALPstarPseudoMagneticFieldLinear}
\end{align}
and
\begin{widetext}
\begin{align}
    \bs{\Upsilon}_q & = \prn{ \bs{\nabla} \Psi^2 } \cos^2\prn{ \bs{k}\cdot\bs{r} - \omega t} - 2 \bs{k} \Psi(r) \cos\prn{ \bs{k}\cdot\bs{r} - \omega t} \sin\prn{ \bs{k}\cdot\bs{r} - \omega t}\,, \nonumber \\
    & = 2 \Psi(r) \pdbyd{\Psi}{r}\cos^2\prn{ \bs{k}\cdot\bs{r} - \omega t} \hat{\bs{r}} - \frac{2 m_a v}{\hbar} \Psi(r) \cos\prn{ \bs{k}\cdot\bs{r} - \omega t} \sin\prn{ \bs{k}\cdot\bs{r} - \omega t} \hat{\bs{k}}\,.
\end{align}
\end{widetext}
There are two generally non-orthogonal components of the signal along $\hat{\bs{k}}$ and $\hat{\bs{r}}$, which change in magnitude as the GNOME sensor passes through the ALP star.
Note that the sensitive axis $\hat{\bs{m}}$ changes in time as Earth rotates, so this can introduce additional time dependence of the signal which is $\propto \hat{\bs{m}} \cdot \bs{\Upsilon}$.
Figure~\ref{Fig:axion-star-signal} shows an example of a simulated signal in a GNOME sensor passing through an ALP star for a particular set of parameters.
Both the direction and magnitude of $\bs{\Upsilon}$ change as the GNOME sensor passes through the ALP star, as well as the direction of $\hat{\bs{m}}$, leading to the somewhat complicated time-dependence.

The GNOME collaboration is currently developing an analysis method to search for ALP stars exhibiting this type of signal using the excess power statistic \cite{anderson2001ExcessPower}.

%%%%%%%%%%%%%%%%%%%%%%%%%%%%%%%%%%%%%%%%%%%%%%%
%%%%%%%%%%%%%%%%%%%%%%%%%%%%%%%%%%%%%%%%%%%%%%%
\subsection{Q-balls}
\label{subsec:Q-balls}
%%%%%%%%%%%%%%%%%%%%%%%%%%%%%%%%%%%%%%%%%%%%%%%
%%%%%%%%%%%%%%%%%%%%%%%%%%%%%%%%%%%%%%%%%%%%%%%

%\DK{Lead writer: Jason/Derek}

Another class of compact composite dark matter objects are \emph{Q-balls} or \emph{Q-stars} \cite{Col85,Lee87,Lee92,Kus01}.
Q-balls are nontopological solitons: bound states of a complex scalar field $\varphi$.
If complex scalar field $\varphi$ obeys global $U(1)$ symmetry, and has a non-zero net charge density Q, the self-interaction of the scalar field makes it energetically favorable for the field to fragment into clumps, called Q-balls \cite{colpi1986boson,nugaev2020review}.
%They can form if $\varphi$ carries a charge from a global $U(1)$ symmetry \cite{nugaev2020review}  and has an appropriate $U(1)$-invariant potential (or due to gravity \cite{colpi1986boson}), and
In some models, Q-balls could comprise the entirety of dark matter \cite{kusenko1998supersymmetric,kusenko1998experimental}.
There are also various theoretical extensions of the Q-ball concept, involving, for example, multiple scalar fields \cite{friedberg1976class} or local $U(1)$ symmetries instead of global \cite{lee1989gauged}.
The distinction between Q-balls and ALP stars is somewhat subtle.
Q-balls, because of the global $U(1)$ symmetry, are true solitons since they are stable due to the conserved charge associated with $\varphi$.
On the other hand, oscillons, our model for ALP stars, represent ``quasi-solitons'' because, while under particular conditions they can be very long-lived, they are technically unstable.
The Q-ball is composed of a complex field, whereas the oscillon is composed of a real field.
For both Q-balls and oscillons, an attractive potential is required for formation.
As we see in the following, the differences between the underlying theoretical models describing oscillons and Q-balls lead to distinct experimental signatures in GNOME.

%\DB{Is there a way to explain this is simple words? What exactly holds a Q-ball together?}
%\DB{I would very much appreciate a plain-language satement of how Q-balls differ from axion stars.}
%\DK{I tried my best to give a decent explanation, see what you think :).}

%%%%%%%%%%%%%%%%%%%%%%%%%%%%%%%%%%%%%%%%%%%%%%%
\subsubsection{Theoretical description}
%%%%%%%%%%%%%%%%%%%%%%%%%%%%%%%%%%%%%%%%%%%%%%%

The complex scalar field of a Q-ball takes the form \cite{Col85}
\begin{align}
    \varphi(\bs{r},t) = \Psi(r) e^{i \omega t}\,,
    \label{eq:Q-ball-field}
\end{align}
where $\omega \lesssim \omega_c$ is the angular oscillation frequency.
The stability of the Q-ball is a consequence of its conserved charge $Q$
\begin{align}
    Q = \frac{\omega}{\hbar^2c^3} \int \left| \varphi(\bs{r},t) \right|^2 d^3\bs{r}~.
\label{Eq:Q-definition}
\end{align}
The necessary conditions for Q-ball formation are that $Q \neq 0$ averaged over the whole space and the existence of a self-interaction potential $U(\Psi)$ possessing at least two distinct minima at $\Psi = 0$ and at $\Psi = \Psi_0$ \cite{Col85,Lee87,Lee92,Kus01}.
In this case, there are regions of space with different vacuum energy values, and the regions where $\Psi = \Psi_0$ can deform but not disappear because of the conserved charge $Q$.
A characteristic feature of the Q-ball is that the potential energy of the field, $U\prn{\Psi}$, is nonzero in the Q-ball's transitional surface region where the field goes from $\Psi = 0$ to $\Psi = \Psi_0$.
Therefore the Q-ball's energy is minimized when its surface area is minimized, leading to a spherical bound-state.
Furthermore, it is energetically favorable for the constituent particles corresponding to the field $\varphi$ (ALPs in our considered case) to remain within the Q-ball in the case where $\omega \lesssim \omega_c$, since ALPs inside the Q-ball have energy $\hbar\omega$ while those outside the Q-star have energy $\hbar\omega_c$.
The values of $\omega^2$ and $\omega_c^2$ are proportional to $\partial^2 U / \partial \Psi^2$ at the respective potential minima inside ($\Psi=\Psi_0$) and outside ($\Psi=0$) the Q-ball, and can thus be different \cite{Col85,Kus01}.
The condition $\omega \lesssim \omega_c$ ensures stability of the Q-star with respect to radiative decay via ALP emission.

%%%%%%%%%%%%%%%%%%%%%%%%%%%%%%%%%%%%%%%%%%%%%%%
\subsubsection{Signal model}
%%%%%%%%%%%%%%%%%%%%%%%%%%%%%%%%%%%%%%%%%%%%%%%

%----------------------------------------------------------------
\begin{figure}
\center
\includegraphics[width=3.35in]{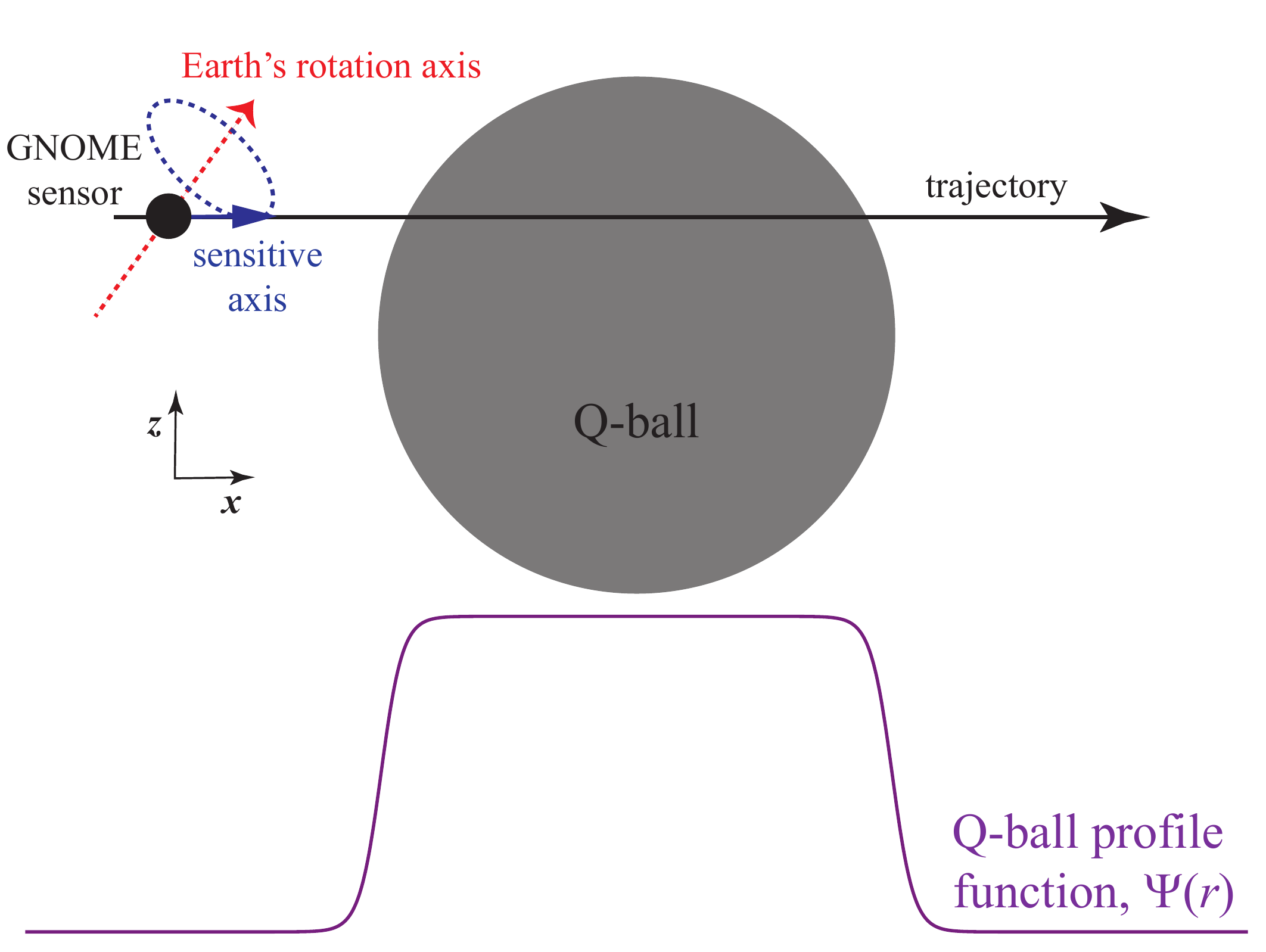}
\includegraphics[width=3.35in]{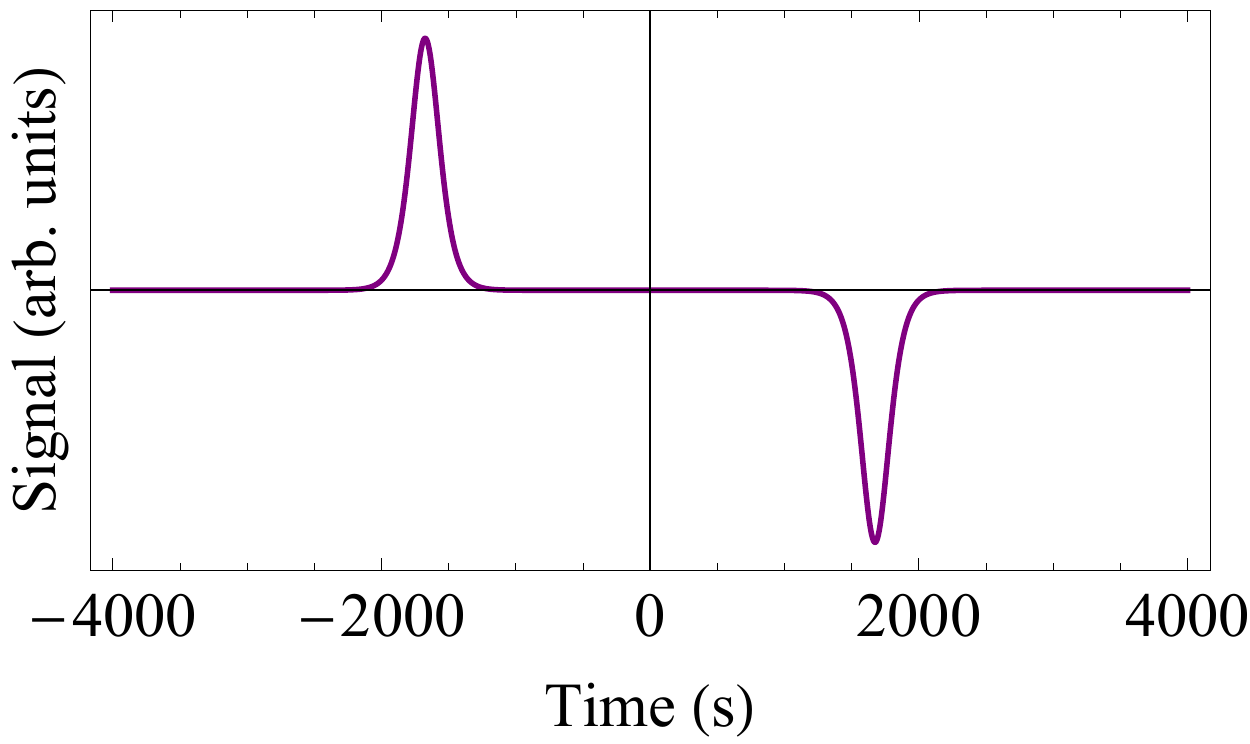}
\caption{Upper diagram: schematic depiction of a GNOME sensor passing through an astronomically large Q-ball (radius $\approx 100 R_E$), indicating with the long black arrow the trajectory of the sensor along $\bs{v}$ (parallel to $\bs{k}$, along $\hat{\bs{x}}$, $v \approx 10^{-3}c$), the initial direction of the GNOME-sensor sensitive axis (blue arrow, along $\hat{\bs{x}}$), and the axis of Earth's rotation (dashed red arrow, $45^\circ$ to the $x$-axis). The purple curve below shows the profile function of the Q-ball. Lower plot: example signal due to the quadratic spin coupling from a Q-ball passage corresponding to the above schematic.}
\label{Fig:Q-ball-signal}
\end{figure}
%----------------------------------------------------------------

The precise form of $\Psi(r)$ depends on details of the potential function, and, although there are special forms of $U(\Psi)$ admitting analytic solutions \cite{rosen1969dilatation,theodorakis2000analytic,gulamov2013analytic}, in general numerical computation is required.
However, there have been a number of analyses, for example Refs.\,\cite{ioannidou2003q,ioannidou2005universality,heeck2021understanding}, that have derived useful approximate radial profile functions that closely correspond to numerical solutions.
For simplicity of mathematical description, here we adopt the form given by Ref.\,\cite{ioannidou2003q},
\begin{align}
    \Psi(r) = \frac{C_0}{\sqrt{ 1 + C_1 \cosh\prn{ \alpha r } }}\,,
\end{align}
where $C_0$, $C_1$, and $\alpha$ are constants determined by fitting to numerical solutions for various potentials and values of $Q$ and $\omega$.
In contrast to the cases of ALP domain walls and ALP stars as described in Secs.~\ref{subsec:domain-walls} and \ref{subsec:axion-stars}, for the complex-valued field $\varphi(r)$ forming Q-balls [Eq.~\eqref{eq:Q-ball-field}], any couplings of spins to $\varphi(r)$ \emph{must} be quadratic in $\varphi(r)$ \cite{kimball2018searching}.
This is required in order to respect the global $U(1)$ symmetry that endows the Q-ball with its charge: from a mathematical point-of-view, interactions must involve products of $\varphi(r)$ and $\varphi(r)^*$ to give a real-valued energy.
Therefore we consider the pseudo-magnetic field $\bs{\Upsilon}_q$ arising from the quadratic interaction,
\begin{align}
        \bs{\Upsilon}_q  & = \bs{\nabla} \abs{ \varphi(r,t) }^2 = \bs{\nabla} \Psi^2(r) \,, \\
        & = \alpha C_0^2 C_1 \frac{\sinh\prn{\alpha r}}{\left[ 1 + C_1 \cosh\prn{\alpha r}\right]^2} \hat{\bs{r}}\,.
\end{align}
Figure~\ref{Fig:Q-ball-signal} shows a schematic example of the signal that would be measured by a GNOME sensor passing through a Q-ball.
In contrast to the ALP star considered in Sec.~\ref{subsec:axion-stars}, the signal of the Q-ball does not exhibit oscillatory behavior, and is in fact more similar to the signature of an ALP domain wall considered in Sec.~\ref{subsec:domain-walls}.
However, there is the notable difference that for a Q-ball there should be two correlated time-separated signals corresponding to entry and exit from the $\Psi = \Psi_0$ region, offering a unique signature pattern that can be exploited for noise rejection.

Note that the signal pattern in GNOME corresponding to a Q-ball encounter (Fig.~\ref{Fig:Q-ball-signal}) closely resembles that from two consecutive domain-wall encounters (Fig.~\ref{Fig:domain-wall-signal}), or a domain-wall encounter with a quadratic coupling in the case of a domain wall that is ``thick'' compared to the size of Earth. This suggests that our data analysis strategies applied to domain-wall searches \cite{masia2020analysis,kim2022machine,afach2021search} can be adapted to the case of Q-balls. Development of an analysis algorithm based on this concept is under way.

%%%%%%%%%%%%%%%%%%%%%%%%%%%%%%%%%%%%%%%%%%%%%%%
%%%%%%%%%%%%%%%%%%%%%%%%%%%%%%%%%%%%%%%%%%%%%%%
\subsection{Dark-matter field fluctuations}
%%%%%%%%%%%%%%%%%%%%%%%%%%%%%%%%%%%%%%%%%%%%%%%
%%%%%%%%%%%%%%%%%%%%%%%%%%%%%%%%%%%%%%%%%%%%%%%

%\DK{Lead writer: Grzegorz/Hector}

%----------------------------------------------------------------
\begin{figure}
\center
\includegraphics[width=3.35in]{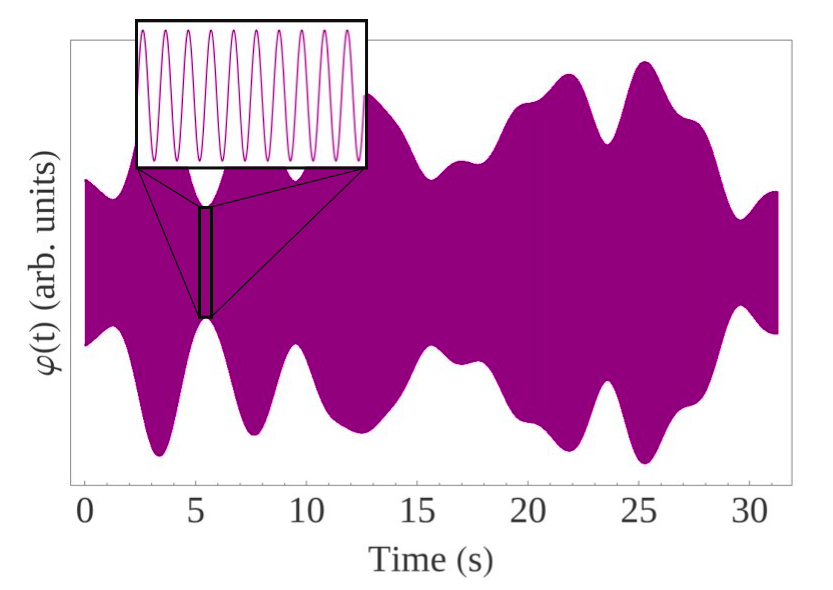}
\caption{Visualisation of the dark matter field oscillating at its Compton frequency with fluctuations of the envelope function.}
\label{Fig:Stochastic-cartoon}
\end{figure}
%----------------------------------------------------------------

Unlike the compact dark matter objects described so far, in this section we consider the more commonly assumed model for UBDM, namely that the UBDM field is spread more evenly throughout the galactic halo volume and not concentrated in large, compact objects.
Due to their ultralight nature and (if they constitute a sizable fraction of the dark matter) their enormous number density, ALPs can be treated as classical plane waves.
In the standard halo model (SHM), ALPs are virialized in the gravitational potential of the Milky Way, and their energy is given by $\hbar \omega \approx m_a c^2 + m_a v^2/2$ where $\omega$ is the observed ALP field oscillation frequency and $v \sim 10^{-3}c$ is the ALP velocity.
Due to the ALPs' randomized velocities, different ALP field modes interfere with one another resulting in a net field that stochastically fluctuates on a characteristic time scale given by the ALP field coherence time $\tau\ts{coh} \sim \hbar/(m_a v^2)$ \cite{foster2018revealing}.
Given that the virial velocity of dark matter in the Milky Way galaxy at the location of our solar system is $\approx 10^{-3}c$ \cite{helmi2002phase,kamionkowski2008galactic}, $\tau\ts{coh} \sim 10^6 \times (2\pi/\omega_c)$.
This means that the ALP field fluctuations occur on a time scale that is a factor of a million slower than the ALP field oscillations.
The stochastically varying properties of the virialized ALP field are well-described by the Rayleigh distribution \cite{centers2021stochastic,derevianko2018detecting}, which, notably, also describes thermal (chaotic) light.
If the coherence length of the virialized ALP field, $\lambda\ts{coh}$, is large compared to the diameter of Earth, GNOME magnetometers could detect common mode fluctuations of the ALP dark matter field.

There are several experiments aiming to detect the ultralight dark-matter field through an observation of a resonant spin coupling to ALPs at their Compton frequency \cite{budker2014proposal,abel2017search,Garcon2017,wu2019search,garcon2019constraints,Aybas2021,roussy2021experimental,Ayb21CASPErE,jiang2021search,bloch2022new,lee2022laboratory,abel2022search}.
Such measurements can also be performed with GNOME and Advanced GNOME sensors.
Given the close analogy between the behavior of virialized ALP fields and thermal light, we can take full advantage of the fact that GNOME is a network of spatially distributed sensors by implementing a detection scheme similar to Hanburry-Brown-and-Twiss intensity interferometry \cite{brown1956correlation} as recently proposed in Ref.~\cite{masia2022intensity}.
The quadratic ALP interaction with spins [Eq.\,\eqref{eq:quadratic-Hamiltonian}] leads to a signal in GNOME magnetometers related to the intensity of the ALP field, $\varphi^2(\bs{r},t)$.
A measurement of the stochastic intensity-like fluctuations of the dark matter field amplitude can be performed using GNOME by correlating time series from distant sensors in order to detect the common-mode signal.
Instead of detecting fast oscillations at $\approx\omega_c$, the slowly varying (at characteristic frequency $\sim 10^{-6}\omega_c$) ALP field envelope function would be measured.
In this way, the probed ALP mass range could be extended to masses several orders of magnitude larger in comparison to resonant measurements performed with the same bandwidth-limited sensors: GNOME's $\sim 100$-Hz-bandwidth sensors could search for ALPs with Compton frequencies up to $\sim 100$~MHz~\cite{masia2022intensity}.

%%%%%%%%%%%%%%%%%%%%%%%%%%%%%%%%%%%%%%%%%%%%%%%
\subsubsection{Theoretical description}
%%%%%%%%%%%%%%%%%%%%%%%%%%%%%%%%%%%%%%%%%%%%%%%

At each point in space, the dark matter field can be described as a superposition of plane waves representing individual modes of the ALP field (i.e., summing over all the individual ALPs composing the field),
\begin{align}
    \varphi(\bs{r},t) = \sum_{n = 1}^{N}   \frac{A}{\sqrt{N}}\cos(\omega_n t- \bs{k}_n\cdot \bs{r} + \theta_n) \, .
    \label{eq:ALPs-superposition}
\end{align}
The amplitude is defined to be $A = \hbar\sqrt{2\rho_{\textrm{dm}}}/(m_ac)$ so that the average ALP field energy density corresponds to the average dark matter density $\rho_{\textrm{dm}}$. The $n$-th ALP has a random velocity $\bs{v}_n$ corresponding to the wave vector $\bs{k}_n = m_{\varphi}\bs{v}_n/\hbar$. Based on the SHM, the probability distribution function of the velocities, $\bs{v}_n$, follows a displaced Maxwell-Boltzmann distribution:
\begin{align}
    f_{\textrm{lab}}(\bs{v}) \approx \frac{1}{\pi^{3/2}v_0^3}\textrm{exp}\left[ -\frac{(\bs{v}-\bs{v}_{\textrm{lab}})^2}{v_0^2} \right] \, ,
\end{align}
where $\left|\bs{v}_{\textrm{lab}}\right| \approx 233 $ km/s is the Sun's velocity with respect to the Galactic rest frame and $v_0\approx 220$ km/s is the velocity dispersion of the ALPs.

The $n$-th oscillation frequency is mostly determined by the Compton frequency $\omega_c = m_{\varphi}c^2/\hbar$, but shifted due to the relativistic Doppler effect for particle waves. For $v_n/c \ll 1$ the effect can be approximated by \cite{gramolin2022spectral}
\begin{align}
    \omega_n = \omega_c \left( 1+\frac{\bs{v}_n^2}{2c^2} \right).
    \label{eq:Stochastic-Doppler-shift}
\end{align}
The fluctuations of the dark matter field in time arise from the random directions, phases, and spread in frequencies of constituent waves. See Fig.~\ref{Fig:Stochastic-cartoon} for a visualisation of the ALP dark matter field amplitude as a function of time.

%----------------------------------------------------------------
\begin{figure}
\center
\includegraphics[width=3.35in]{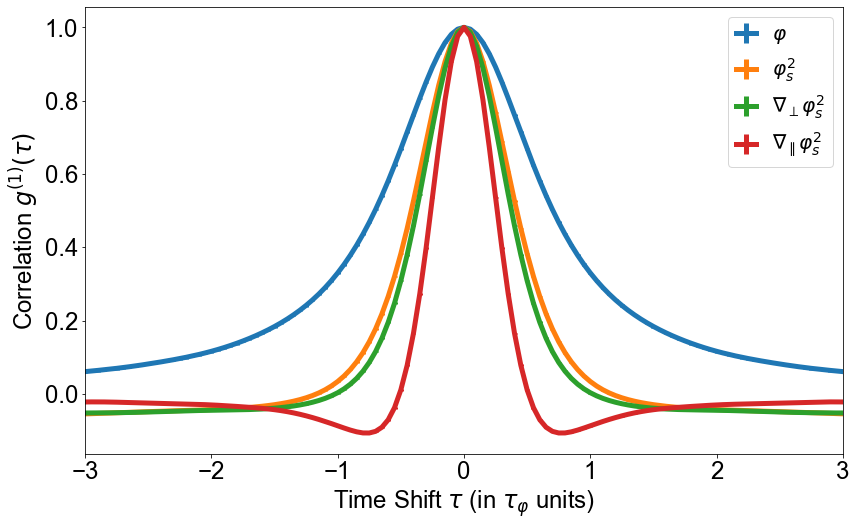}
\caption{Auto-correlation of a simulated ALP dark matter field $\varphi$, the low-frequency component of its square $\varphi_s^2$ associated with stochastic fluctuations, and two gradient components perpendicular and parallel to $\bs{v}_{\mathrm{lab}}$: namely $\bs{\nabla}_{\perp}\varphi_s^2$ and $\bs{\nabla}_{\parallel}\varphi_s^2$, where the subscript $s$ denotes the low-frequency component associated with stochastic fluctuations.}
\label{Fig:Stochastic-correlation}
\end{figure}
%----------------------------------------------------------------

%%%%%%%%%%%%%%%%%%%%%%%%%%%%%%%%%%%%%%%%%%%%%%%
\subsubsection{Signal model}
%%%%%%%%%%%%%%%%%%%%%%%%%%%%%%%%%%%%%%%%%%%%%%%

Assuming the quadratic coupling Hamiltonian [Eq.~\eqref{eq:quadratic-Hamiltonian}], the pseudo-magnetic field is given by $\bs{\Upsilon}_q = \bs{\nabla} \varphi^2$ and the signal observed in a GNOME sensor is $\propto \hat{\bs{m}} \cdot \bs{\Upsilon}_q$.
As the detectors have limited bandwidth $\Delta \omega$, for the case where $\Delta \omega \ll \omega_c$ the fast oscillations at $2\omega_c$ are not observable in the measurement.
The measured pseudo-magnetic field can be approximated by the near-dc component
\begin{align}
    \bs{\Upsilon}_q = \bs{\nabla} \varphi^2 \approx \frac{A^2}{2N}\sum_{n,m = 1}^{N}\bs{k}_{nm} \sin(\omega_{nm}t - \bs{k}_{nm}\cdot \bs{r}+\theta_{nm}),
\end{align}
where $\omega_{nm} = \omega_n-\omega_m$, $\bs{k}_{nm} = \bs{k}_n-\bs{k}_m$, and $\theta_{nm} = \theta_n - \theta_m$.
Such a signal can be well characterized by two parameters: coherence time and amplitude.

For a solitary magnetometer or comagnetometer, a UBDM signature would be difficult to distinguish from comparably large background noise.
Our strategy for analysis of GNOME data, presently underway, is to measure the cross-correlation between data measured at different stations.
For stations with aligned sensitive axes and separations $\lesssim \lambda\ts{coh}$ and time delays $\lesssim \tau\ts{coh}$, UBDM quadratically coupled to atomic spins could generate a common-mode signal that might be distinguishable from uncorrelated background noise for sufficiently strong interactions.
Figure~\ref{Fig:Stochastic-correlation} shows the auto-correlation function of simulated ALP signals, which is equivalent to the cross-correlation between stations with aligned sensitive axes within a coherence patch (ignoring any contribution from uncorrelated noise).
The width of the correlation feature is $\approx \tau\ts{coh}$ in relative time shift units.
This technique of intensity interferometry will enable a search for the quadratic coupling to ALPs over a mass range $\sim 10^{-14}$ eV -- $10^{-9}$ eV using GNOME \cite{masia2022intensity}.

%%%%%%%%%%%%%%%%%%%%%%%%%%%%%%%%%%%%%%%%%%%%%%%
%%%%%%%%%%%%%%%%%%%%%%%%%%%%%%%%%%%%%%%%%%%%%%%
\subsection{Solar axion halo}
\label{subsec:solar-axion-halo}
%%%%%%%%%%%%%%%%%%%%%%%%%%%%%%%%%%%%%%%%%%%%%%%
%%%%%%%%%%%%%%%%%%%%%%%%%%%%%%%%%%%%%%%%%%%%%%%

%\DK{Lead writer: Derek}

\begin{figure}
\center
\includegraphics[width=3.35in]{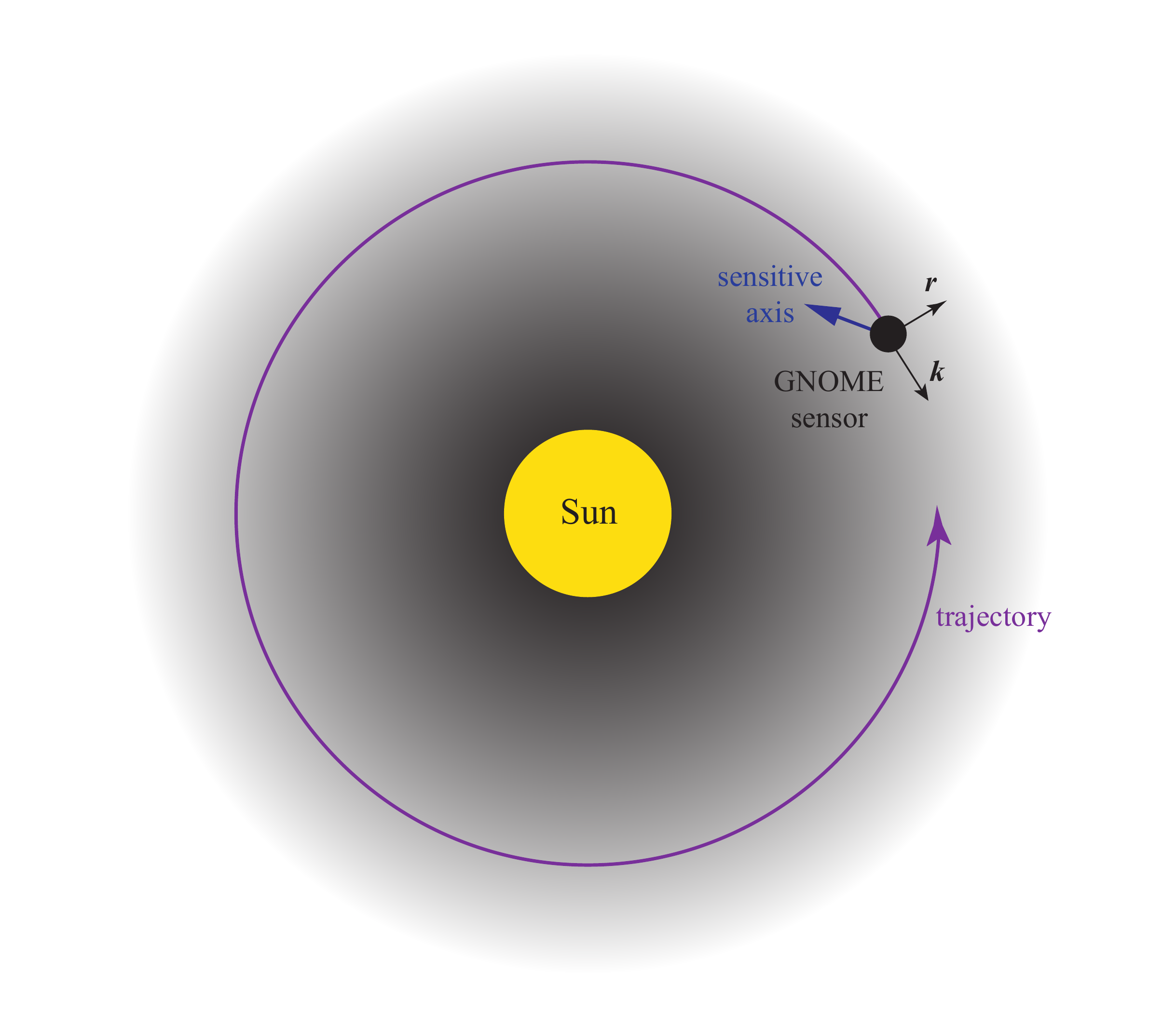}
\includegraphics[width=3.15in]{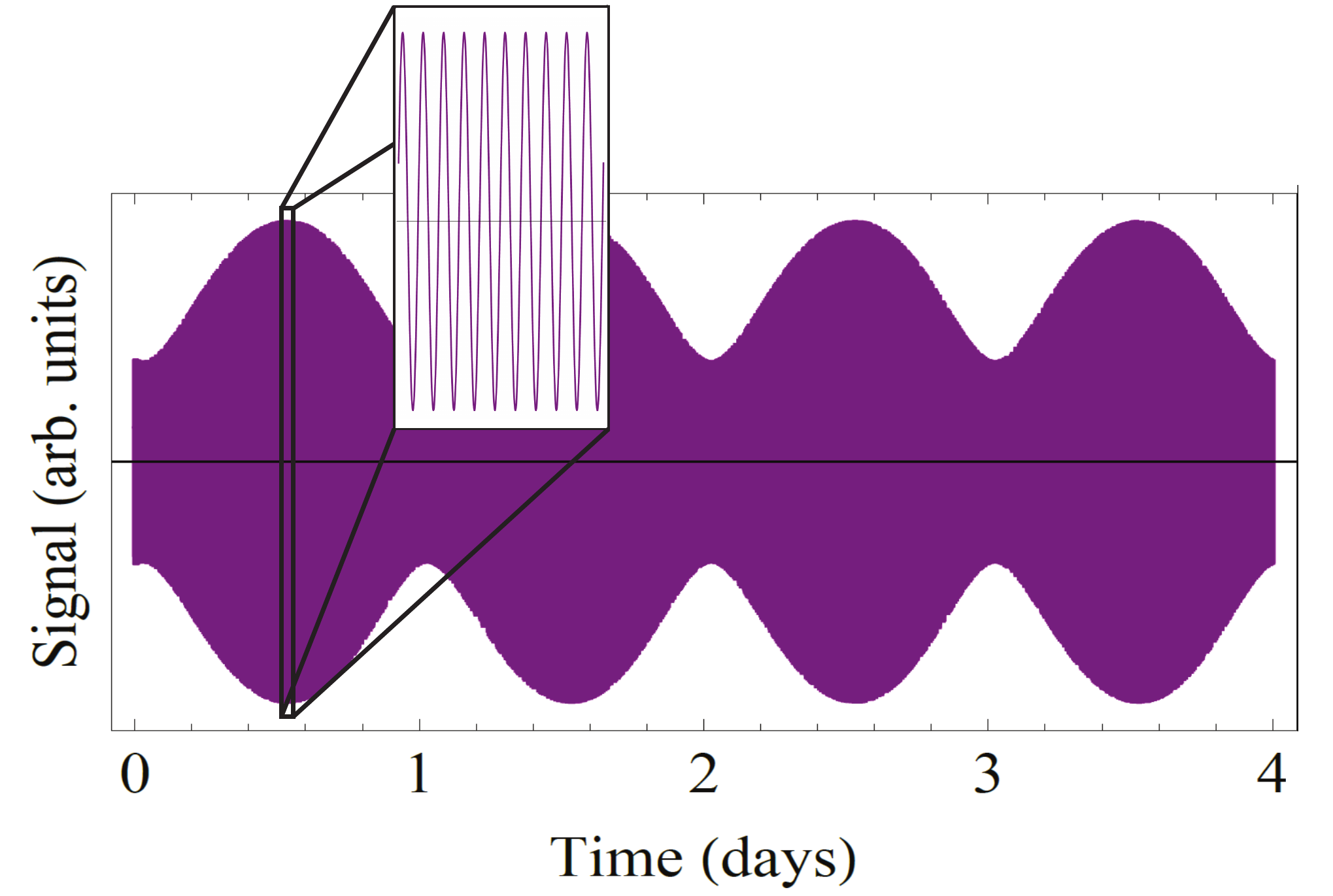}
\caption{Upper diagram: Schematic diagram of Earth moving through a solar ALP halo. The pseudo-magnetic field associated with a solar ALP halo coupling to atomic spins has both a radial component in the $-\hat{\bs{r}}$ direction due to the spatial gradient and a transverse component due to the ``ALP  wind'' directed along the ALP halo's relative velocity with respect to the lab frame, $\hat{\bs{k}}$. Lower plot: example signal due to the linear spin coupling. Signal is shown over four days, exhibiting an envelope function due to the daily modulation of the signal due to the rotation of the GNOME sensor's sensitive axis with respect to the pseudo-magnetic field $\bs{\Upsilon}$. Inset shows the fast oscillation at the Compton frequency.}
\label{Fig:solar-ALP-halo-signal}
\end{figure}

Another theoretical possibility, closely related to the axion star model described in Sec.~\ref{subsec:axion-stars}, is that tidal shearing forces may enable astrophysical bodies such as Earth and the Sun to capture UBDM in their gravitational fields and form a local halo \cite{banerjee2020relaxion,grote2019novel,banerjee2020searching,vermeulen2021direct,tretiak2022improved}, see Fig.~\ref{Fig:solar-ALP-halo-signal}.
In these models, the Earth or Sun effectively acts as a seed in the formation process of the axion star.
If such a process occurs, there would be a substantial overdensity of the ALP field near these bodies as compared to the average dark matter density as discussed in Refs.\,\cite{banerjee2020relaxion,banerjee2020searching}.
Possible mechanisms for capture of dark matter by dense astrophysical bodies have been studied, for example, in Refs.\,\cite{xu2008dark,khriplovich2009capture,khriplovich2011capture,khriplovich2011capture2,brito2015accretion,brito2016interaction}, and continue to be actively investigated.

%----------------------------------------------------------------
%----------------------------------------------------------------

%%%%%%%%%%%%%%%%%%%%%%%%%%%%%%%%%%%%%%%%%%%%%%%
\subsubsection{Theoretical description}
%%%%%%%%%%%%%%%%%%%%%%%%%%%%%%%%%%%%%%%%%%%%%%%

Here we assume the model of a gravitationally bound solar ALP halo described in Refs.~\cite{banerjee2020relaxion,banerjee2020searching}, and also assume that the halo is non-rotating (i.e., at rest with respect to the Sun).
For a solar halo, the ALP field amplitude at the position of Earth exponentially decays over a characteristic length scale given by
\begin{align}
    R_\star \approx \frac{\hbar^2}{G_N M_\odot m_a^2}\,,
    \label{eq:Rstar}
\end{align}
where $G_N$ is Newton's gravitational constant and $M_\odot$ is the Sun's mass.
For the solar ALP halo to extend to the position of Earth, we require $R_\star \gtrsim 1 \, \textrm{AU}$.
This imposes the requirement that $m_a c^2 \lesssim 10^{-14}~{\rm eV}$.
The ALP field oscillates at $\approx \omega_c$ with a coherence time $\tau\ts{coh} \gtrsim m_a R_\star^2 / \hbar$.
For $m_a c^2 \lesssim 10^{-14}~{\rm eV}$, the coherence time is longer than $\sim 10^7$~s; therefore, we can treat oscillations of the ALP field as effectively single frequency (i.e., monochromatic).
Thus the ALP field can be described as \cite{banerjee2020relaxion,banerjee2020searching}:
\begin{align}
    \varphi(\bs{r},t) \approx \varphi_0 \cos\prn{ \omega_c t - \bs{k}\cdot\bs{r} + \theta } e^{ - r / R_\star }\,,
    \label{eq:solar-halo-ALP-field}
\end{align}
where $\varphi_0$ is a constant determined by the overall energy density in the solar ALP halo and $\theta$ is a random phase, constant over the coherence time and coherence length.
For simplicity, in this example we set $\theta = 0$.

%%%%%%%%%%%%%%%%%%%%%%%%%%%%%%%%%%%%%%%%%%%%%%%
\subsubsection{Signal model}
%%%%%%%%%%%%%%%%%%%%%%%%%%%%%%%%%%%%%%%%%%%%%%%

The exotic pseudo-magnetic field due to a solar ALP halo for the linear spin interaction is given by
\begin{widetext}
\begin{align}
    \bs{\Upsilon}_l = \bs{\nabla}\varphi(\bs{r},t) = \varphi_0 e^{ - r / R_\star } \sbrk{ \bs{k} \sin\prn{ \omega_c t - \bs{k}\cdot\bs{r}} -  \frac{\hat{\bs{r}}}{R_\star} \cos\prn{ \omega_c t - \bs{k}\cdot\bs{r}} }\,,
\end{align}
and for the quadratic spin interaction is given by
\begin{align}
    \bs{\Upsilon}_q = \bs{\nabla}\varphi^2(\bs{r},t) = \varphi_0^2 e^{ - 2r / R_\star } \sbrk{ \bs{k} \sin\prn{ 2\omega_c t - 2\bs{k}\cdot\bs{r}} - \frac{2\hat{\bs{r}}}{R_\star} \cos^2\prn{ \omega_c t - \bs{k}\cdot\bs{r}} }~.
\end{align}
\end{widetext}
We note that in both cases there are two components of $\bs{\Upsilon}$ in the lab frame \cite{banerjee2020searching}: (1) a radial component from the spatial ALP gradient directed toward the Sun's position and (2) a transverse component along $\bs{k}$ due to the ALP wind interaction \cite{graham2018spin,stadnik2014axion}.
Because the relative velocity of Earth with respect to the Sun is dominated by its orbital motion ($\sim 100$ times faster than the velocity component due to Earth's rotation about its axis), it turns out that all GNOME sensors have approximately the same $\bs{k}$ and thus see the same field.
The amplitudes of the radial and transverse components of $\bs{\Upsilon}$ are relatively constant in time, so the signal observed in a particular GNOME sensor has a predictable daily modulation due to the time dependence of $\hat{\bs{r}}\cdot\hat{\bs{m}}$ and $\hat{\bs{k}}\cdot\hat{\bs{m}}$ caused by the rotation of Earth.
An example of the time dependence of such a signal is shown in the lower plot of Fig.~\ref{Fig:solar-ALP-halo-signal}.

Analysis of GNOME data to search for a solar ALP halo is in progress. The strategy is to search for a persistent single-frequency signal in cross-correlation data between stations, and then use the predicted daily modulation to distinguish between a signal from a solar ALP halo and systematic backgrounds.

%%%%%%%%%%%%%%%%%%%%%%%%%%%%%%%%%%%%%%%%%%%%%%%
%%%%%%%%%%%%%%%%%%%%%%%%%%%%%%%%%%%%%%%%%%%%%%%
\subsection{Exotic Low-mass Fields (ELFs) emitted from black hole mergers}
%%%%%%%%%%%%%%%%%%%%%%%%%%%%%%%%%%%%%%%%%%%%%%%
%%%%%%%%%%%%%%%%%%%%%%%%%%%%%%%%%%%%%%%%%%%%%%%

%\DK{Lead writer: Ibrahim/Sami}
In Ref.~\cite{dailey2021quantum}, a subset of the present authors considered a possibility that bursts of ultrarelativistic scalar fields can be produced in the course of some high energy astrophysical event such as a supernova explosion, a binary black hole merger, or in conjunction with a fast radio burst via some as yet unknown coherent process\footnote{Note that due to the energy-time uncertainty relation, if the high-energy astrophysical event generating the ELF has duration $\tau_0$, the corresponding spread in energies of the emitted ELF wave packet is $\gtrsim \hbar/\tau_0$. Thus the spread in frequencies at the source is $\Delta \omega_0 \gtrsim 2\pi/\tau_0$. Our discussion here assumes coherent production, which requires $\Delta \omega_0 \ll \omega$. To interpret results in terms of the model presented here and in Ref.~\cite{dailey2021quantum}, we must restrict considerations to the regime where the relativistic energy $\varepsilon = \hbar \omega$ of the ELF satisfies $\varepsilon \gg \hbar/\tau_0$, as noted in Ref.~\cite{stadnik2021comment}.} (see, for example, Ref.~\cite{Eby:2021ece}).
%{\color{blue}YS: I think there's some confusion here. $\tau_0 \gtrsim \hbar / \Delta \omega_0$ is a consequence of the time-energy uncertainty relation. Coherent production requires $\Delta \omega_0 \ll \omega$. The original paper \cite{dailey2021quantum} didn't discuss this, so it makes sense to cite the relevant comment arXiv:2111.14351 here.}
%{\color{blue}YS: I recall from my comment in arXiv:2111.14351 that there was an inconsistency in \cite{dailey2021quantum}, like in figure 3 therein, for some combinations of ELF masses and burst durations as to whether the ELF burst could be produced coherently. The time-energy uncertainty relation dictates that there is a minimum spread in ELF energies produced by a burst of finite duration, which can prevent coherent production at smaller ELF masses (we'd instead get incoherent production).}
These ``exotic low-mass fields'' (ELFs) would thus have large mode occupation numbers, enabling their treatment as classical waves with oscillation frequency given by $\hbar \omega = \sqrt{ m^2 c^4 + p^2 c^2} \approx p c$. Such ELFs can be searched for with GNOME.

%%%%%%%%%%%%%%%%%%%%%%%%%%%%%%%%%%%%%%%%%%%%%%%
\subsubsection{Theoretical description}
%%%%%%%%%%%%%%%%%%%%%%%%%%%%%%%%%%%%%%%%%%%%%%%

Consider a scenario where some energy $\Delta E$ is radiated isotropically in the form of a Gaussian ELF burst with duration $\tau_0$ at the source.
The Klein-Gordon equation for $\varphi(r)$ can be solved to obtain
\begin{align}
\varphi(t) \approx & \frac{A_0}{R}\sqrt{\frac{\tau_0}{\tau}}\exp \prn{-\frac{(t-t_s)^2}{2 \tau^2}} \nonumber\\
                  & \times \cos\prn{ \omega_0 (t-t_s) - \frac{\omega_0}{4\delta t}(t-t_s)^2 }\,,
\label{Eq:DispersionModel}
\end{align}
where $\omega_0$ is the central frequency of the wave packet, $\tau$ the duration of the pulse as measured by sensors on Earth, $t_s = R/v_g$ represents the transit time of the ELF pulse from a source at distance $R$ from Earth, and $v_g$ the group velocity.
The delay time between an electromagnetic or gravitational wave trigger and an ELF pulse is $\delta t = R/v_g - R/c$.
In terms of the total energy released, the amplitude $A_0$  is given by
\begin{equation}
    A_0  \approx \frac{1}{\pi^{1/4}}\prn{\frac{1}{\omega_0}\sqrt{\frac{c\Delta E}{2\pi \tau_0}}}\, .
    \label{Eq:AmplitudeGaussian}
\end{equation}
Equations \eqref{Eq:DispersionModel} and \eqref{Eq:AmplitudeGaussian} describe a Gaussian wavepacket which disperses with a chirp rate
\begin{equation}
\frac{d\omega(t)}{dt} = - \frac{1}{\tau_0 \tau} = - \frac{\omega_0}{2 \delta t} \,.
\label{Eq:Slope}
\end{equation}
Notably, the chirp rate is constrained by the central frequency and the delay from the electromagnetic or gravitational wave trigger.

%----------------------------------------------------------------
\begin{figure}
\center
\includegraphics[width=3.35in]{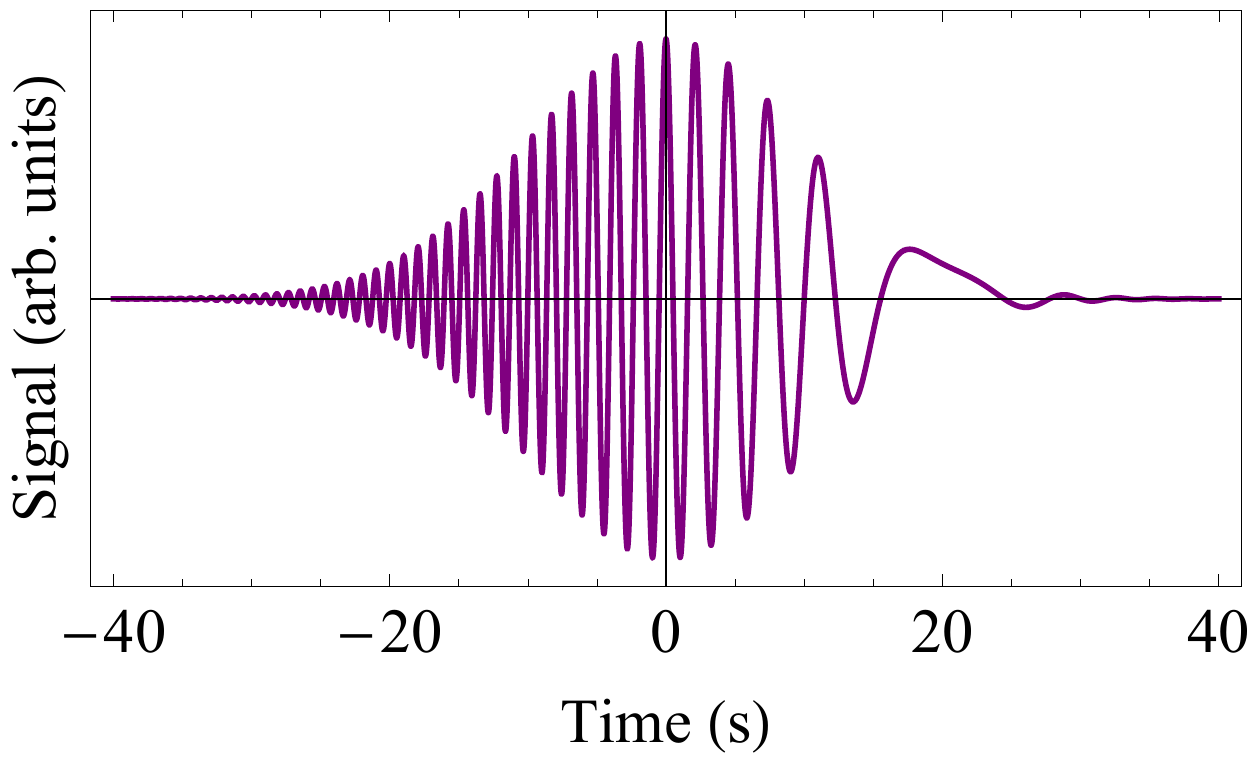}
\caption{Example signal of an ELF measured by a GNOME sensor based on Eq.~\eqref{Eq:DispersionModel}, with $\omega_0 =2\pi\times (0.5~{\rm Hz})$, $\tau = 10$~s, and $\delta t = 10$~s.}
\label{Fig:ELF-signal}
\end{figure}
%----------------------------------------------------------------

%%%%%%%%%%%%%%%%%%%%%%%%%%%%%%%%%%%%%%%%%%%%%%%
\subsubsection{Signal model}
%%%%%%%%%%%%%%%%%%%%%%%%%%%%%%%%%%%%%%%%%%%%%%%

As a consequence of the couplings described in Eqs.~\eqref{eq:linear-interaction-Lagrangian} to \eqref{eq:quadratic-Hamiltonian}, the existence of scalar fields with a form such as that described by Eq.~\eqref{Eq:DispersionModel} would lead to pseudo-magnetic fields in GNOME due to interaction terms proportional to $\bs{S} \cdot \bs{\nabla} \varphi(\bs{r},t) $ and $\bs{S} \cdot \bs{\nabla} [\varphi(\bs{r},t)]^2$.
A signal due to an ELF event (Fig.~\ref{Fig:ELF-signal}) would therefore: (1) originate from the same source as the electromagnetic or gravitational wave trigger; (2) have a spatial pattern in the GNOME network given by the projection of the ELF velocity vector on the axes of the sensors in the network; and (3) have a characteristic chirp pattern given by Eq.\,\eqref{Eq:Slope}.
Using these characteristics, we have designed an analysis algorithm to search for ELF signals in GNOME data.
Measurements from the GNOME sensors are projected into the time-frequency plane with a resolution chosen to match the expected signal dispersion.
We then construct a network-wide test statistic and search for signals consistent with ELFs in the GNOME data in intervals of time shortly after reported binary black hole merger and fast radio burst events.

\begin{figure}[ht!]
\centering
\includegraphics[width=3.35in]{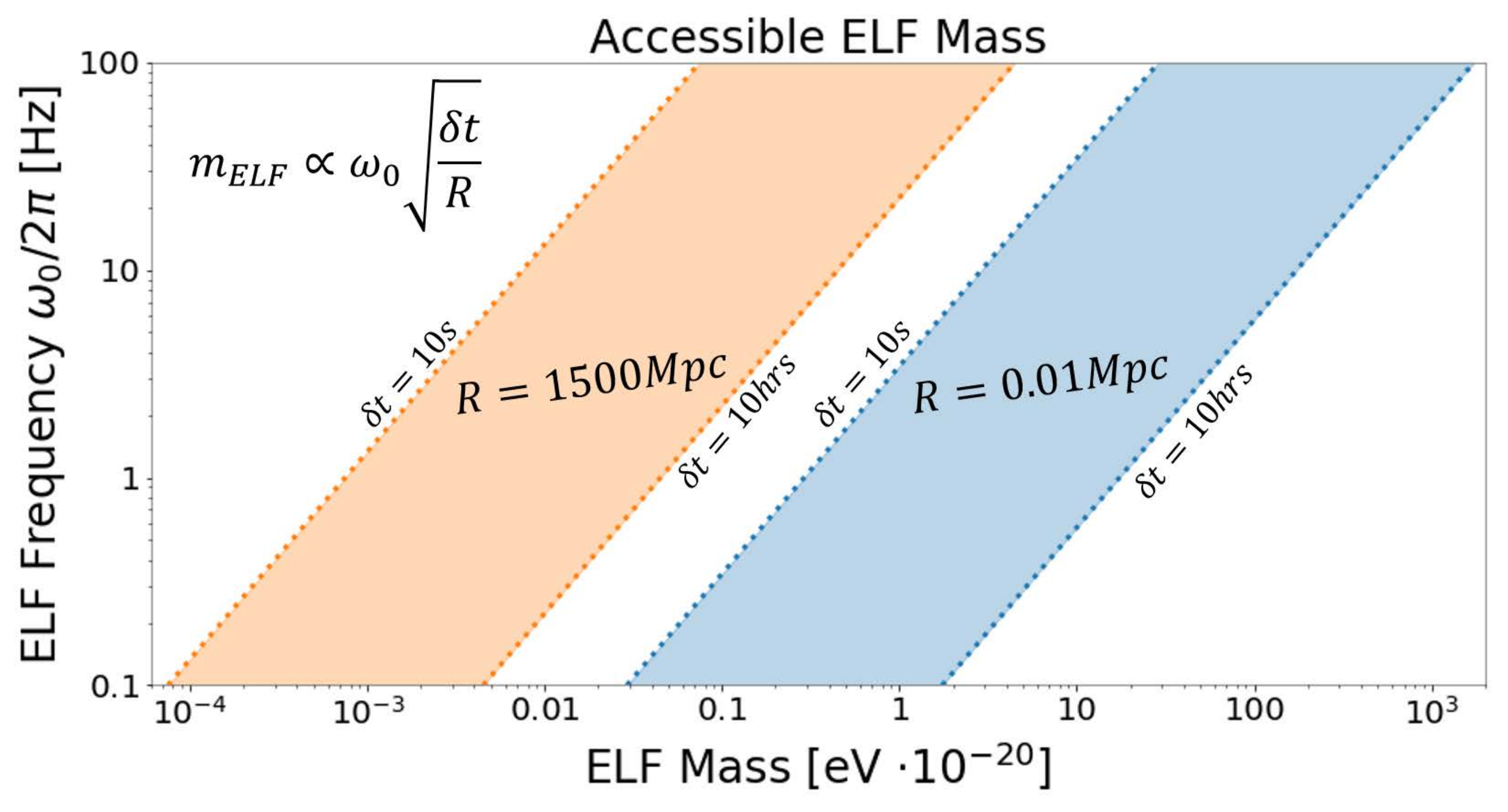}
\caption{Accessible masses for ELFs from sources located at distances R = 1500 Mpc and R = 0.01 Mpc. The left and right edges of the bands are set by the delay times $10 \, \rm{s} < \delta t < 10 \, \rm{hr}$.}
\label{fig:ParSpace}
\end{figure}

We show in Fig.~\ref{fig:ParSpace} the range of masses that are accessible by GNOME given the bandwidth of the magnetometers, assuming an observation time of 10 hours after the trigger.
Thus GNOME can act as a ``telescope'' to detect ELF signals from sources that generate ALP bursts of sufficient intensity.

%%%%%%%%%%%%%%%%%%%%%%%%%%%%%%%%%%%%%%%%%%%%%%%
%%%%%%%%%%%%%%%%%%%%%%%%%%%%%%%%%%%%%%%%%%%%%%%
\subsection{Other potential candidates}
%%%%%%%%%%%%%%%%%%%%%%%%%%%%%%%%%%%%%%%%%%%%%%%
%%%%%%%%%%%%%%%%%%%%%%%%%%%%%%%%%%%%%%%%%%%%%%%

There are many other exotic physics scenarios that can potentially be searched for using data from the GNOME experiment, demonstrating the versatility of the general approach.
For example, as already noted, it may be that there are relatively strong interactions between particles in the dark sector.
Such interactions could cause dark matter particles to coalesce into large composite ``blobs'' \cite{grabowska2018detecting}.
In contrast to the axion star, Q-ball, and solar axion-halo scenarios already considered above, which are composite states of ultralight particles, there could be dark matter blobs formed from heavy fermions or bosons that exert long-range forces mediated by other fundamental constituents of the dark sector \cite{grabowska2018detecting,baum2022searching}.
Other possible search targets for the GNOME experiment are ALP waves emanating from black hole superradiance \cite{baryakhtar2021black} or from the collapse of axion stars \cite{Eby:2021ece}.
In related work, unshielded magnetometer networks have been used to search for signals from dark (or hidden) photons \cite{fedderke2021earth,fedderke2021search} and axions \cite{arza2022earth}.

%\DB{There are other things that can be mentioned here. For example, the use of networks to search for AQN: Dmitry Budker, Victor V. Flambaum, Xunyu Liang, and Ariel Zhitnitsky, Axion Quark Nuggets and how a Global Network can discover them, Phys. Rev. D 101, 043012 (2020), arXiv:1909.09475
%This particular paper should only be mentioned briefly as it is tangential to GNOME.}

%\DB{However (!) I suggest a separate Section on how advanced GNOME sensors can be be used to search for oscillating signals that are not transient but oscillating (what Arne is doing now).}

%%%%%%%%%%%%%%%%%%%%%%%%%%%%%%%%%%%%%%%%%%%%%%%
%\subsubsection{Dark matter blobs}
%%%%%%%%%%%%%%%%%%%%%%%%%%%%%%%%%%%%%%%%%%%%%%%

%\DK{Lead writer: Deniz}

%%%%%%%%%%%%%%%%%%%%%%%%%%%%%%%%%%%%%%%%%%%%%%%
%\subsubsection{Black hole superradiance}
%%%%%%%%%%%%%%%%%%%%%%%%%%%%%%%%%%%%%%%%%%%%%%%

%\DK{Lead writer: Derek}

%%%%%%%%%%%%%%%%%%%%%%%%%%%%%%%%%%%%%%%%%%%%%%%
%\subsubsection{Signals from spin-1 bosons}
%%%%%%%%%%%%%%%%%%%%%%%%%%%%%%%%%%%%%%%%%%%%%%%

%\DK{Lead writer: ??}

%%%%%%%%%%%%%%%%%%%%%%%%%%%%%%%%%%%%%%%%%%%%%%%
%\subsubsection{Gravitational focusing of UBDM streams}
%%%%%%%%%%%%%%%%%%%%%%%%%%%%%%%%%%%%%%%%%%%%%%%

%\DK{Lead writer: Ophir/Abaz}

\begin{figure}
\center
\includegraphics[width=3.35in]{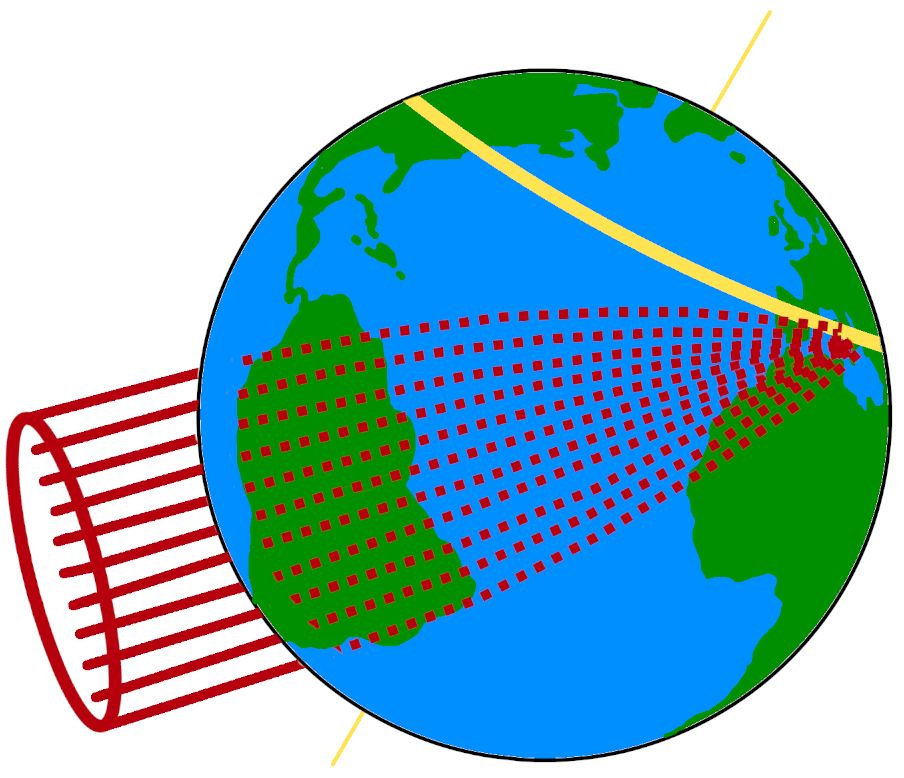}
\caption{Example of a low velocity UBDM stream (red) gravitationally focused into a small region on Earth's surface, the trajectories are approximately following predictions made in Ref.~\cite{Sofue:2020}. As Earth rotates around its axis, the focus follows a path (yellow) along a fixed latitude.}
\label{Fig:GF-cartoon}
\end{figure}

Yet another possible search target for GNOME are UBDM streams, in particular those gravitationally focused by other bodies in the solar system.
Cosmological simulations indicate that up to $10^{12}$ small-velocity-dispersion (compared to that in the SHM) streams of DM may be present in the vicinity of the solar system~\cite{Vogelsberger:2011}.
The deflection experienced by these particles depends on the mass distribution of the source of gravitational focusing, is independent of the DM particle mass, and is inversely proportional to the square of its velocity.
For UBDM, its wave nature has to be considered for a full description of the spectral signatures due to gravitational focusing~\cite{Kim:2021yyo}
Generally, for deflection of particles by the gravitational potential of the Earth, efficient focusing on Earth's surface (where GNOME operates) occurs for particles with velocities of approximately 10 to 20\,km/s~\cite{Sofue:2020, gfstream:2022}.
In this case, flux enhancements of up to $10^9$ are predicted.
Gravitational deflection by other objects in the solar system, such as Jupiter, can also generate flux enhancements on Earth.
This requires larger stream velocities of approximately $10^{-3}$\,c and results in smaller enhancements of only $10^{6}$~\cite{patla:2014}.
Again, the networked approach of GNOME offers several advantages when searching for such gravitationally focused streams.
Namely, the small size of such stream foci, on the order of a few km for deflection in Earth's gravitational potential and much less for that of Jupiter, leads to transient signals lasting only about 10\,s.
This can be difficult to distinguish from noise when only a single detector is employed.
For long streams, there is a chance of the detecting the enhanced DM flux daily with one or more detectors over the course of several weeks.
Obviously, the likelihood of encountering the high flux regions increases with the number of sensors in a network.
A network also allows for precisely determining the spatial direction of the stream, which can help to better understand the origins of such streams when detected, although this can be a complex endeavor as upstream sources of gravitational deflection also have to be considered.
%\AK{Changed the sentence above to reflect what the reference quoted is about and asked below from Dima/Hendrik}
%\DB{This is actually a comment from Hendrik: I'm not sure about the following sentence, where does this idea come from? This strongly depends on the direction of the stream, right? Which is biased slightly in the direction of our velocity through the halo. And then the latitude where the focussing occurs still depends on, for example, the velocity of the stream.}
%\AK{For low velocity streams the distribution of focus regions is isotrotropic (only low velocity streams focus near the Earth's surface) so all geographical locations have equal probability to encounter these high flux regions.   }
Furthermore, sensors at approximately the same latitudes could produce correlated signals from the same focused-DM source (Fig. ~\ref{Fig:GF-cartoon}). The geographical locations of GNOME sensors make it viable for such correlations.

%----------------------------------------------------------------
%----------------------------------------------------------------
%\DB{Hendrik Bekker and Akaash are working hard on a paper significantly extending earlier focusing analyses. I suggest inviting Hendrik to contribute. OR: Done.}

%%%%%%%%%%%%%%%%%%%%%%%%%%%%%%%%%%%%%%%%%%%%%%%
%\subsubsection{GNOME as CASPEr}

%\DK{Lead writer: Dima}\\

%%%%%%%%%%%%%%%%%%%%%%%%%%%%%%%%%%%%%%%%%%%%%%%

While GNOME is primarily a tool for detection of transient ``new-physics'' signals, its sister experiment, the Cosmic Axion Spin-Precession Experiment (CASPEr) \cite{budker2014proposal,Garcon2017,Ayb21CASPErE}, is searching for oscillating (but not necessarily transient) effects of galactic dark matter and dark matter concentrated in the Sun- and Earth-bound halos.
CASPEr operates simultaneously in two locations, Boston University (USA) and the Helmholtz Institute, Johannes Gutenberg University, Mainz (Germany).
The central idea of CASPEr is that it is a nuclear magnetic resonance (NMR) experiment, where the role of the transverse oscillating magnetic field ``$\bs{B}_1$'' is relegated to the ambient Axion/ALP field either through its coupling to gluons (CASPEr-electric) or via its gradient interactions [see Eqs.\,\eqref{eq:linear-Hamiltonian} and \eqref{eq:quadratic-Hamiltonian}; CASPEr-gradient].
While CASPEr will reach up to Compton frequencies $\approx$600\,MHz, it has already ventured into the much lower-frequency range \cite{wu2019search,garcon2019constraints}, from near DC to several hertz.
The low-frequency range has also been explored by various other experiments \cite{abel2017search,jiang2021search,bloch2022new,lisanti2021stochastic,abel2022search}.

The Advanced-GNOME comagnetometers \cite{padniuk_response_2022,klinger2022polarization} are ideally suited to contribute to ``CASPEr-like'' searches in the Compton frequency range from near-DC to several hundred hertz, with the sensitivity that is estimated to be competitive to the best previous searches.
The results of the first experiments in this direction may be expected already at the early stages of Advanced GNOME, as a search in this mode does not require the full network capabilities.
However, as with GNOME, we expect that the network will be immediately useful in vetoing spurious signals from various laboratory inferences, which will likely be different in different laboratories.
Moreover, the network offers interesting possibilities of correlating the stochastic behavior of the signal in different stations that may offer an additional tool for discriminating the new-physics signals.

%%%%%%%%%%%%%%%%%%%%%%%%%%%%%%%%%%%%%%%%%%%%%%%
%%%%%%%%%%%%%%%%%%%%%%%%%%%%%%%%%%%%%%%%%%%%%%%
\section{Conclusion and outlook}
%%%%%%%%%%%%%%%%%%%%%%%%%%%%%%%%%%%%%%%%%%%%%%%
%%%%%%%%%%%%%%%%%%%%%%%%%%%%%%%%%%%%%%%%%%%%%%%

%\DK{Lead writer: Dima}

The idea of GNOME is to carry out synchronous measurements using optical atomic magnetometers and comagnetometers operating within magnetically shielded environments in distant locations.
The GNOME experiment searches for a class of signals different from that probed by most other experiments, namely transient and stochastic effects that could arise from ALP fields of astrophysical origin passing through Earth during a finite time.
By analyzing the correlation between the signals from multiple, geographically separated sensors, it is possible to probe a wide variety of exotic physics scenarios.

When the idea of GNOME was conceived, ALP domain walls were the sole targets of the search \cite{Pos13,pustelny2013global}.
As the collaboration expanded and matured, we gradually realized that there are many more interesting physics scenarios that can be investigated with this tool, as, we hope, we have been able to convince the reader of this ``GNOME science-case'' article.
Extrapolating from the first 10 years of the GNOME collaboration's work, we may only expect the science case to continue its expansion, hopefully leading to a discovery of ``new'' physics.

The advent of GNOME was soon followed by the establishment of networks of atomic clocks \cite{Der14} introduced to search for scalar UBDM whose effect is to produce apparent oscillation of fundamental constants, rather than causing spin precession.
At the same time, it was realized that certain UBDM candidates like the relaxion \cite{graham2015cosmological,banerjee2019coherent} may have mixed intrinsic parity and displaying both the scalar and pseudoscalar properties.
This provides a motivation for building hybrid-sensor networks (for example, (co)magnetometer-clock networks) to cross-correlate the signals and enhance the background-suppression capabilities.
We expect this cross-network approach to become commonplace in future work.
Moreover, GNOME is planning future science runs to be synchronized with gravitational-wave detector networks such as LIGO/VIRGO in order to carry out ``exotic'' multimessenger astronomy \cite{dailey2021quantum}.

It appears that the science case for search networks such as GNOME is mostly limited by our current scientific horizons and there are, in fact, many other interesting and unexpected signals we could be looking for, had we better imagination.
But could we perhaps look for such signals before we realize what they are?
It appears that modern machine-learning approaches indeed offer us such an exciting opportunity.
It might work along the following lines: a machine-learning system is trained on the data from GNOME (and/or an expanded hybrid-sensor network) and establishes the ``normal'' dynamic state of the network.
It is then further trained on a variety of new-physics scenarios.
As a result of the training that can continue for the lifetime of the experiment, the artificial intelligence (AI) system would be able to recognize unusual events or patterns on the network.
While this may look like a remote prospect, one should recognize that the task at hand is not too different from that of detection of unidentified threats, which is routinely done at present by automated security systems.

In conclusion of our overview of the ``GNOME science'', we mention that, apart from fundamental physics, magnetometer networks may also be useful in more ``down-to-Earth'' applications such as, for example, the study of magnetic signal patterns in urban environments and what they reveal about cities \cite{Bowen2019Geo,Dumont2022JAP}.

\section*{Acknowledgements}

The authors are grateful to Konstantin Zioutas for bringing the gravitational focusing effects to the attention of the GNOME collaboration.
The GNOME collaboration thanks the Mainz Institute for Theoretical Physics (MITP) of the Cluster of Excellence ``Precision Physics, Fundamental Interactions, and Structure of Matter'' (PRISMA+ EXC 2118/1) for its hospitality and its partial support during the completion of this work.
This work was additionally supported by the U.S. National Science Foundation under grants PHY-1307507, PHY-1707875, PHY-1707803, PHY-2110370,  PHY-2110388, and PHY-2207546; by the German Research Foundation (DFG) under grant no. 439720477 and within the German Excellence Strategy (Project ID 39083149); by COST (European Cooperation in Science and Technology);
ZDG acknowledges institutional funding provided by the Institute of Physics Belgrade through a grant by the Ministry of Science, Technological Development and Innovations of the Republic of Serbia.
The work of SP and GL is supported by the National Science Centre of Poland within the Opus program (grant 2019/34/E/ST2/00440) and in part by the Excellence Initiative -- Research University of the Jagiellonian University in Krak\'ow.
The work of YVS was supported by the Australian Research Council under the Discovery Early Career Researcher Award DE210101593.
The work of AD was supported in part by the Heising-Simons Foundation. The work of JE was supported by the World Premier International Research Center Initiative (WPI), MEXT, Japan and by the JSPS KAKENHI Grant Numbers 21H05451 and 21K20366.
RF and DFJK acknowledge support by the NSF-BSF grant No. 2016635.

%{\color{blue}YS: At the moment, only a few of the references have (embedded) hyperlinks, while most don't - it would be great to have consistency one way or the other. }
%\TS{I have put DOI numbers.}

\bibliography{GNOME}

%%%%%%%%%%%%%%%%%%%%%%%%%%%%%%%%%%%%%%%%%%%%%%%%%%%%%%%%%%%%%%%%%%%%%%%%%%%%%
%%%%%%%%%%%%%%% Notes on references %%%%%%%%%%%%%%%%%%%%%%%%%%%%%%%%%%%%%%%%%
%%%%%%%%%%%%%%%%%%%%%%%%%%%%%%%%%%%%%%%%%%%%%%%%%%%%%%%%%%%%%%%%%%%%%%%%%%%%%

\end{document}